\documentclass{aastex}
\usepackage{graphicx,epsfig,emulateapj5} 
\usepackage{amsmath}

\newcommand{\etal}{{et al.~}}

\newcommand{\bq}{\begin{equation}}
\newcommand{\eq}{\end{equation}}

\def\gtsim{\lower.5ex\hbox{$\buSildrel > \over\sim$}}
\def\ltsim{\lower.5ex\hbox{$\buildrel < \over\sim$}}
\def\arcsec{^{\prime\prime}}

\def\farcs{\hbox{$.\!\!^{\prime\prime}$}}

\def\apjl{ApJL}
\def\apj{ApJ}
\def\apjs{ApJS}
\def\mnras{MNRAS}
\def\araa{ARAA}
\def\aj{AJ}
\def\aap{A\&A}
\def\aaps{A\&A Suppl.}

\slugcomment{Accepted by the Astrophysical Journal}

\shortauthors{Heiderman \etal}
\shorttitle{Interacting Galaxies in the A901/902 Supercluster}

\begin{document}

\title {Interacting Galaxies in the A901/902 Supercluster with STAGES}


\author{Amanda Heiderman\altaffilmark{1}, Shardha
Jogee\altaffilmark{1}, Irina Marinova\altaffilmark{1}, Eelco van
Kampen\altaffilmark{9,17}, Marco Barden\altaffilmark{9}, Chien
Y. Peng\altaffilmark{5,6}, Catherine Heymans\altaffilmark{12}, Meghan
E. Gray\altaffilmark{2}, Eric F. Bell\altaffilmark{4}, David
Bacon\altaffilmark{7}, Michael Balogh\altaffilmark{8}, Fabio
D. Barazza\altaffilmark{3}, Asmus B\"ohm\altaffilmark{10,9}, John
A.R. Caldwell\altaffilmark{16}, Boris H\"au\ss ler\altaffilmark{2},
Knud Jahnke\altaffilmark{4}, Kyle Lane\altaffilmark{2}, Daniel H.
McIntosh\altaffilmark{13}, Klaus Meisenheimer\altaffilmark{4},
Sebastian F. S\'anchez\altaffilmark{11}, Rachel S.
Somerville\altaffilmark{4}, Andy Taylor\altaffilmark{12}, Lutz
Wisotzki\altaffilmark{10}, Christian Wolf\altaffilmark{14}, $\&$
Xianzhong Zheng\altaffilmark{15} }

 \altaffiltext{1}{Department of Astronomy, University of Texas at
Austin, 1 University Station C1400, Austin, TX 78712-0259}
\altaffiltext{2}{School of Physics and Astronomy, The University of
Nottingham, University Park, Nottingham NG7 2RD, UK}
\altaffiltext{3}{Laboratoire d'Astrophysique, \'Ecole
Polytechnique F\'ed\'eralede Lausanne (EPFL), Observatoire, CH-1290
Sauverny, Switzerland} 
\altaffiltext{4}{Max-Planck-Institut f\"{u}r
Astronomie, K\"{o}nigstuhl 17, D-69117, Heidelberg, Germany}
\altaffiltext{5}{NRC Herzberg Institute of Astrophysics, 5071 West
  Saanich Road, Victoria, V9E 2E7, Canada}
\altaffiltext{6}{Space Telescope Science Institute, 3700 San Martin
  Drive, Baltimore, MD 21218, USA}
\altaffiltext{7}{Institute of Cosmology and Gravitation, Dennis Sciama
Building, University of Portsmouth, Dennis Sciama Building,
Portsmouth, PO1 3FX, UK}
\altaffiltext{8}{Department of Physics and Astronomy, University Of
Waterloo, Waterloo, Ontario, N2L 3G1, Canada}
\altaffiltext{9}{Institute for Astro- and Particle Physics,
University of Innsbruck, Technikerstr. 25/8, A-6020 Innsbruck,
Austria}
\altaffiltext{10}{Astrophysikalisches Insitut Potsdam, An der
Sternwarte 16, D-14482 Potsdam, Germany} 
\altaffiltext{11}{Centro Hispano Aleman de Calar Alto, C/Jesus Durban
  Remon 2-2, E-04004 Almeria, Spain}
\altaffiltext{12}{The Scottish Universities Physics Alliance (SUPA),
  Institute for Astronomy, University of Edinburgh, Blackford Hill,
  Edinburgh, EH9 3HJ, UK}
\altaffiltext{13}{Department of Physics, 5110 Rockhill Road,
  University of Missouri-Kansas City, Kansas City, MO 64110, USA}
\altaffiltext{14}{Department of Physics, Denys Wilkinson Building,
  University of Oxford, Keble Road, Oxford, OX1 3RH, UK}
\altaffiltext{15}{Purple Mountain Observatory, National Astronomical
  Observatories, Chinese Academy of Sciences, Nanjing 210008, PR
  China}
\altaffiltext{16}{University of Texas, McDonald Observatory, Fort
Davis, TX 79734, USA}
\altaffiltext{17}{ESO, Karl-Schwarschild-Str. 2, D-85748, Garching bei
M\"unchen, Germany}

\begin{abstract}
We present a study of galaxy mergers and the influence of environment
in the Abell 901/902 supercluster at $z\sim$~0.165, based on 893
bright ($R_{\rm Vega} \le$~24) intermediate mass ($M_{*} \geq 10^{9}
M_{\sun}$) galaxies.  We use $HST$ ACS F606W data from the Space
Telescope A901/902 Galaxy Evolution Survey (STAGES), COMBO-17,
$Spitzer$ 24$\micron$, and $XMM$-$Newton$ X-ray data.  Our analysis
utilizes both a physically driven visual classification system, and
quantitative CAS parameters to identify systems which show evidence of
a recent or ongoing merger of mass ratio $ >$~1/10 (i.e., major and
minor mergers).  Our results are: (1)~After visual classification and
minimizing the contamination from false projection pairs, we find that
the merger fraction $f_{\rm merge}$ is 0.023$\pm$0.007.  The estimated
fractions of likely major mergers, likely minor mergers, and ambiguous
cases are 0.01$\pm$0.004, 0.006$\pm$0.003, and 0.007$\pm$0.003,
respectively.  (2)~All the mergers lie outside the cluster core of
radius $R <$~0.25 Mpc: the lack of mergers in the core is likely due
to the large galaxy velocity dispersion in the core. The mergers,
instead, populate the region (0.25 Mpc $< R\leq$~2 Mpc) between the
core and the cluster outskirt. In this region, the estimated frequency
of mergers is similar to those seen at typical group overdensities in
N-body simulations of accreting groups in the A901/902 clusters. This
suggests ongoing growth of the clusters via accretion of group and
field galaxies.  (3)~We compare our observed merger fraction with
those reported in other clusters and groups out to $z\sim$~0.4.
Existing data points on the merger fraction for $L \leq L^{*}$
galaxies in clusters allow for a wide spectrum of scenarios, ranging
from no evolution to evolution by a factor of $\sim$5 over
$z\sim$~0.17 to 0.4.  (4)~In A901/902, the fraction of mergers, which
lie on the blue cloud is 80$\%\pm 18\%$ (16/20) versus 34$\%\pm7\%$ or
(294/866) for non-interacting galaxies, implying that interacting
galaxies are preferentially blue.  (5)~The average SFR, based on UV or
a combination of UV+IR data, is enhanced by a factor of $\sim$1.5 to 2
in mergers compared to non-interacting galaxies.  However, mergers in
the A901/902 clusters contribute only a small fraction (between 10\%
and 15\%) of the total SFR density, while the rest of the SFR density
comes from non-interacting galaxies.
\end{abstract}


\section{Introduction}\label{sintro}

Understanding how galaxies evolve in various environments (field,
groups, and clusters), and as a function of redshift is a key step
toward developing a coherent picture of galaxy evolution.  Present-day
cluster and field galaxies differ due to several factors, which are
often grouped under the umbrella of `nature' versus `nurture'.
First, in cold dark matter (CDM) cosmogonies, the first galaxies
formed and evolved early in cluster cores, as the higher initial
overdensities led to faster gravitational collapse and more rapid
mergers of proto-galaxies (e.g., Cole et \etal 2000; Steinmetz \&
Navarro 2002).  Second, in the context of the bottom-up CDM assembly
paradigm, the outer parts of clusters and superclusters, grow at late
times via mergers, smooth accretion, and discrete accretion of groups
and field galaxies.  This idea is supported by observational studies
(e.g., Zabludoff \& Franx 1993; Abraham \etal 1996a; Balogh, Navarro
\& Morris 2000), which suggest that clusters continuously grew by the
accretion of groups.  Third, the dominant physical processes
affecting galaxies differ in cluster and field environments due to the
different galaxy number density, galaxy velocity dispersion, and
intracluster medium (ICM).  Among these processes are close
galaxy-galaxy interactions, such as strong tidal interactions and
mergers (e.g., Barnes 1992; Moore \etal 1998) and galaxy harassment
(e.g., Moore \etal 1996), which stems from the cumulative effect of
weak interactions.  Furthermore, in clusters where the hot ICM makes
up as much as 15\% of the total mass, galaxy-ICM interactions, such as
ram pressure stripping (Gunn \& Gott 1972; Larson \etal 1980; Quilis
\etal 2000; Balogh, Navarro \& Morris 2000), can play an important
role in removing the diffuse gas from galaxies. The tidal field of the
cluster potential may also play a relevant role in the dynamical
evolution of cluster galaxies (Gnedin 2003).

Systematic studies of the differences between cluster, group, and
field galaxies at different redshifts are needed to shed light on the
relative importance of these various processes in their respective
environments.  Several differences have been observed between galaxies
in the field and those in the rich cluster environment, but the
physical drivers behind these variations are still under
investigation.  At $z\sim$~0, the relative percent of massive early
type (E+S0) galaxies to spirals rises from (10\%+10\%:80\%) in the
field to (40\%+50\%:10\%) in the cores of very rich clusters, leading
to the so-called morphology density relation (Dressler 1980; Dressler
\etal 1997).  However, recent Sloan Digital Sky Survey (SDSS) studies
suggest that masses and star formation (SF) histories of galaxies are
more closely related to environmental physical processes rather than
their structural properties (Blanton \etal 2005).  The SF histories of
galaxies depend on both luminosity (Cole et al. 2001) and environment
(Diaferio \etal 2001; Koopmann \& Kenney 2004).  The fraction of blue
galaxies in clusters appears to rise with redshift, an effect known as
the Butcher Oemler effect (Butcher \& Oemler 1978; Margoniner \etal
2001; de Propis \etal 2003).  There is also evidence that SF in bright
($M_{\rm v} < $~-18) cluster galaxies is suppressed compared to field
galaxies (e.g., Balogh \etal 1998, 1999), for reasons that are not
well understood.

Galaxy interactions and mergers have been proposed as a mechanism for
the change in galaxy populations in clusters from that of the field
(e.g., Toomre \& Toomre 1972; Lavery \& Henry 1988; Lavery Pierce, \&
McClure 1993).  There have been various studies on the properties of
galaxies (e.g., Dressler 1980, Postman \& Geller 1984, Giovanelli,
Haynes, \& Chincarini, 1986, Kennicutt 1983; Gavazzi \& Jaffe 1985,
Whitmore \etal 1993) and of galaxy interactions and mergers (e.g.,
Lavery \& Henry 1988; Lavery, Pierce, \& Mclure 1992; Zepf 1993;
Dressler \etal 1994; Couch \etal 1998; van Dokkum \etal 1998, 1999;
Tran \etal 2005, 2008) in different environments.  Some of these
studies suggest that galaxy interactions and mergers may play a role
in morphological transformations of galaxies in clusters, but there
have been few systematic studies of galaxy mergers and interactions in
clusters, based on high resolution $HST$ images as well as $Spitzer$
24$\micron$, and X-ray images.

In this paper we present a study of the frequency, distribution,
color, and SF properties of galaxy mergers in the A901/902
supercluster at $z\sim$0.165. We use $HST$ ACS F606W data taken as
part of the Space Telescope A901/902 Galaxy Evolution Survey (STAGES;
Gray \etal 2009), along with ground-based COMBO-17 imaging data (Wolf
\etal 2004), $Spitzer$ 24$\micron$ data (Bell \etal 2005, 2007),
$XMM$-$Newton$ X-ray data (Gilmour \etal 2007), and dark matter (DM) mass
measurements from weak lensing (Heymans \etal 2008).
With a resolution of $0.1\arcsec$ or $\sim$280 pc at $z=0.165$, the $HST$
images allow for the identification of merger signatures such as
double nuclei, arcs, shells, tails, tidal debris, and accreting
satellites.  The COMBO-17 survey (Wolf \etal 2004) provides accurate
spectrophotometric redshifts down to $R_{\rm Vega}$ of 24 and stellar
masses (Borch \etal 2006).  The $Spitzer$ 24$\micron$ data (Bell \etal
2005, 2007) probe the obscured SF, while X-ray maps (Gilmour \etal
2007; Gray \etal, 2010 in preparation) provide information of how the ICM density
changes throughout the cluster.

We present the data and sample selection in $\S$Section~\ref{sdatas}.  In
$\S$Section~\ref{smvc} and $\S$Section~\ref{smcas1}, we describe the two different
methods that we use to identify galaxy mergers: a physically motivated
classification system that uses visual morphologies, stellar masses,
and spectrophotometric redshifts, and a method based on CAS asymmetry
$A$ and clumpiness $S$ parameters (Conselice \etal 2000). In
$\S$Section~\ref{sinte1} and~\ref{scas1}, we explore the frequency of galaxy
mergers in A901/902 based on these two methods and present one of the
first systematic comparisons to date between CAS-based and visual
classification results in clusters.  We set a lower limit on the
fraction of major mergers (those with mass ratio $M_{1}/M_{2} \ge
1/4$).  In $\S$Section~\ref{sinte2}, we examine the distribution of mergers
in the A901/902 supercluster as a function of clustocentric radius,
galaxy number density, local galaxy surface density, relative ICM
density, and local DM mass surface density.  In $\S$Section~\ref{sinte3}, we
compare our results on the fraction and distribution of mergers to
expectations based on analytical estimates and simulations of mergers
in different environments.  In $\S$Section~\ref{scomp1}, we compare our
results on galaxy mergers in the A901/902 supercluster to groups and
clusters at different redshifts out to $z\sim$~0.8.  We investigate
the fraction of mergers and non-interacting galaxies on the blue cloud
and red sample as a function of clustocentric radius in
$\S$Section~\ref{scolor1}.  Finally in $\S$Section~\ref{ssfr1}, we compare the star
formation rate (SFR) of mergers and non-interacting galaxies in the
A901/902 clusters. The results of this work are summarized in
$\S$Section~\ref{ssumm}.  In this paper, we assume a flat cosmology with
$\Omega_{\rm m}$ = 1 $-\Omega_{\lambda}$ = 0.3 and H$_{0}$~=~70 km
s$^{-1}$ Mpc$^{-1}$ throughout this paper.

\section{The A901/902 Supercluster: Dataset and Sample Selection}\label{sdatas}

The A901/902 supercluster is composed of three clusters: A901a, A901b,
and A902, and related groups (Gray \etal 2002; Heymans \etal 2008).
This study utilizes data from the STAGES survey (Gray \etal 2009),
which provides high resolution F606W $Hubble$ $Space$ $Telescope$
($HST$) Advanced Camera for Surveys (ACS) images over a $0.5^{\circ}
\times 0.5^{\circ}$ field.  Additional multi-wavelength data include
$XMM$-$Newton$ (Gilmour \etal 2007), $Spitzer$ 24$\micron$ data (Bell
\etal 2005, 2007), and ground based COMBO-17 imaging data (Wolf \etal
2004).

Stellar masses are taken from Borch \etal (2006).  They were derived
by fitting the 17-band COMBO spectral energy distributions (SEDs) with
a library of template SEDs, which were constructed using the PEGASE
stellar population synthesis model, assuming different SF histories
and a Kroupa (Kroupa, Tout, \& Gilmore 1993) initial mass function
(IMF) in the mass regime 0.1--120~$M_{\sun}$.  Such stellar masses are
consistent within 10\% with masses estimated using a Kroupa (2001) or
Chabrier (2003) IMF\footnote{We adopt a Chabrier (2003) IMF when
exploring the contribution of mergers to the SFR in $\S$Section~\ref{ssfr1}}.

Accurate spectrophotometric redshifts down to $R_{\rm Vega}\sim$24
and spectral energy distributions, based on 5 broad bands ($UBVRI$)
and 12 medium band filters, are available from the COMBO-17 project
(Wolf \etal 2004).  The 1-sigma redshift error ($\sigma_{z(R)}$) modeled
from spectroscopic redshifts that are available for a subset of 420
galaxies, scales as:

\bq\label{eqz1} \case{\sigma_{z(R)}}{(1 + z)}= 0.005 \times \sqrt{1 +
10^{0.6(R - 20.5)}} \eq where $R$ is the apparent $R$-band magnitude
of the galaxy measured in a 1.5$\arcsec$ aperture.

The reduction of the ACS images and the selection of the A901/902
supercluster sample are described in detail in Gray et al. (2009), and
we only provide a summary here.  STAGES A901/902 object detection was
automated using source extractor (SExtractor; Bertin \& Arnouts 1996)
on the ACS F606W images, yielding $\sim$12,500 galaxies with matching
COMBO-17 counterparts down to $R_{\rm Vega} \sim$24. A subset of these
galaxies were considered to be supercluster members based on their
spectrophotometric redshifts.  The spectrophotometric redshifts of
cluster members were assumed to follow a Gaussian distribution with a
half-width ($\Delta z$) related to $\sigma_{z(R)}$. A galaxy is
considered to be a cluster member if its spectrophotometric redshift
$z$ lies in the range [0.17$-\Delta z$, 0.17+$\Delta z$], where

\bq\label{eqz2} 
\Delta z(R)=\sqrt{0.015^{2} +(1.65 \sigma_{z(R)})^{2}} 
\eq

The resulting dependence of the half-width $\Delta z$ on $R$ ensures a
completeness level of $>$ 90\% at all magnitudes for $R_{\rm Vega}
\leq$ 24.  The field contamination (i.e., the fraction of galaxies
that are field members) for this sample is estimated by assuming that
the average number count of field galaxies as a function of magnitude
and redshift is consistent with the trends determined from regions
lying outside of the cluster. The amount of field contamination rises
strongly at fainter magnitude due to the increase of $\sigma_{z(R)}$
at fainter R, and typical values are 20\% at $R_{\rm Vega}$ = 21.65
and $\sim$70\% at $R_{\rm Vega}$ = 24 (see Figures 13 and 14 of Gray
\etal 2009).

The top panel of Figure~\ref{flumf} shows the distribution of apparent
$R$--band magnitude, absolute magnitude $M_{\rm V}$, and stellar mass
for the sample of 1990 galaxies with $R_{\rm Vega} \le$24.  We obtain
our working sample of 893 galaxies by applying a further cut of $M_{*}
\geq10^{9}$ $M_{\sun}$ for the following reasons.  First, for $M_{*}
\geq10^{9}$ $M_{\sun}$, the sample is complete in stellar mass on the
red cloud and blue sequence (Borch et al. 2006), while at lower masses
incompleteness sets in.  Second, as shown in the middle panel of
Figure~\ref{flumf}, most galaxies with $R_{\rm Vega} \le$24 and $M_{*}
\geq10^{9}$ $M_{\sun}$ actually have apparent magnitudes of $R_{\rm
Vega} \le$21.7, where the field contamination is $\le$20\% (see
Figures 13 and 14 of Gray \etal 2009).  Third, most galaxies with
$M_{*} \geq10^{9}$ $M_{\sun}$ have a luminosity $M_{\rm V} \le -18$
(Figure~\ref{flumf}), where the cluster luminosity function (Binggeli,
Sandage, \& Tammann 1988) is typically dominated by E to Sd galaxies,
rather than small dwarf galaxies. The former galaxies are large and
many of their morphological features (e.g., disks, bars, bulges,
spiral arms, double nuclei, arcs, shells, tails, tidal debris, and
accreting satellites) can be resolved by the STAGES ACS F606W images
whose effective point spread function (PSF) of $\sim$0.1$\arcsec$
correspond to $\sim$280 pc at $z$ = 0.165.

\section{Methodology  and Analysis}\label{smetho}

\subsection{Galaxy Mergers in this study}\label{smdef}

Before outlining the methods ($\S$Section~\ref{smvc} and $\S$Section~\ref{smcas1})
that we use to identify mergers, we briefly discuss the types of
mergers that are of interest to our study, their morphological
signatures, and their typical post-merger products.

According to simulations, galaxy mergers of mass ratio
$M_{1}/M_{2}>$~1/10 tend to have a significant impact on galaxy
evolution.  They include major mergers, which are defined to be those
with a mass ratio of 1/4 $< M_{1}/M_{2} \le$ 1/1, as well as minor
mergers with 1/10 $< M_{1}/M_{2} \le$ 1/4.  Simulations show that
major mergers of stellar systems typically destroy the outer disks,
transforming them via violent relaxation into systems with a steep
central surface brightness profile, associated with a high S\'ersic
index $n$ (typically $n>$ 2.5; Hopkins \etal 2009).  These remnants
include ellipticals and spheroids with a de Vaucouleurs-type stellar
profile ($n$~=~4) (e.g., Negroponte \& White 1983; Barnes \& Hernquist
1991; Mihos \& Hernquist 1996; Struck 1997; Naab \& Burkert 2001).  In
gas-rich major mergers, a disky component can form or survive inside
the resulting spheroidal component (Hopkins \etal 2009), producing a
remnant whose overall profile is less steep than a de Vaucouleurs one,
with $n<4$.  Irrespective of the details of the remnants, the above
simulations suggest that ongoing/recent major mergers are associated
with arcs, shells, ripples, tidal tails, large tidal debris, extremely
asymmetric light distributions, double nuclei inside a common body,
and tidal bridges/envelopes of light linking systems of similar mass.

Conversely, minor mergers involving a spiral galaxy and a smaller
satellite of mass ratio 1/10 $< M_{1}/M_{2} \le$ 1/4, will not destroy
the outer disk of the larger companion (e.g., Hernquist \& Mihos 1995;
Smith \etal 1999; Jogee \etal 1999).  Typically, the smaller companion
sinks via dynamical friction, may excite bars, spirals, and other
non-axisymmetric perturbations in the disk of the larger galaxy, and
leads to tidal heating, arcs, shells, ripples, tidal tails, tidal
debris, warps, offset rings, and highly asymmetric light distributions
(e.g., Quinn \etal 1993; Hernquist \& Mihos 1995; Mihos \etal 1995;
Quinn, Hernquist, \& Fullagar 1993; Smith 1997; Jogee \etal 1999;
Jogee 2006 and references therein).

In this paper, we identify merging systems using two independent
methods, which make use in different ways of the aforementioned
morphological signatures seen in simulations. The first is a
physically motivated classification system ($\S$Section~\ref{smvc}), which is
similar to that defined in Jogee \etal (2008, 2009), and is based on
visual morphologies, stellar masses, and spectrophotometric redshifts.
The second method ($\S$Section~\ref{scas1}) uses the CAS merger criterion ($A
>$~0.35 and $A > S$), which is based on quantitative asymmetry ($A$),
and clumpiness ($S$) parameters derived using the CAS code (Conselice
\etal 2000).

\subsection{Visual classification of mergers and non-interacting 
galaxies}\label{smvc}

We visually classify the ACS F606W images of galaxies in the sample of
893 bright ($R_{\rm Vega} \le$~24) intermediate mass ($M_{*} \geq
10^{9}$) galaxies.  This sample is complete for $M_{*} \geq10^{9}$
$M_{\sun}$ (see $\S$Section~\ref{sdatas}).  In $\S$Section~\ref{sinte1}, we will
discuss whether further mass cuts should be applied in order to ensure
the completeness of major and minor mergers, but for now we consider
all the systems in the sample of 893 galaxies.  A small fraction
(below 1\%) of the sample could not be classified due to image
defects, low signal to noise, and a highly compact appearance.  As
illustrated in Figure~\ref{fvcsch}, the remaining galaxies are
classified into 5 main visual classes: Mergers of types 1, 2a, and 2b,
and Non-interacting systems of type Symmetric and Irr-1.

The visual class `Mergers' is assigned to systems with evidence of a
recent or ongoing merger of mass ratio $>$~1/10. The mergers are
subdivided into 3 groups called Type 1, Type 2a, and Type 2b, because
different criteria are used to identify these three types of mergers
and different techniques are used to separate them into major and
minor mergers.  Non-interacting systems are subdivided into the visual
classes of Symmetric and Irr-1.  The results of the visual
classification are illustrated in Tables~\ref{tvc1}-~\ref{tmer},and
Figures~\ref{fmer}-~\ref{fexam3}.  Below are details of the 5 main
visual classes.

\vspace{3 mm} {\bf (A) Mergers of type 1: } Mergers of type 1 are
systems, which appear as a single distorted remnant in the ACS F606W
image (with effective PSF $\sim$$0\farcs1$ or $\sim$280 pc at $z$ =
0.165), and which host morphological distortions similar to those
produced in the afore-mentioned simulations of mergers of mass ratio
$>$1/10. Thus, mergers of type 1 likely represent the very advanced
phases of such a merger, at a point where the 2 progenitor galaxies
have coalesced into a single merger remnant.  Among the sample of
$M_{*} \geq 10^{9} M_{\sun}$ systems, we find 13 mergers of type
1. Their properties are shown in Table~\ref{tmer} and
Figure~\ref{fmer}. They show morphological distortions such as tidal
tails (Figure~\ref{fmer}, Cases 10, 11), shells, ripples, warps,
asymmetric tidal debris and distortions (e.g., Cases 5, 6, 7, 9), or a
`train-wreck' morphology (Cases 7, 10, 11, 12).

For mergers of type 1, a single redshift and stellar mass are
available. Thus, the evidence for a merger of mass ratio $>$~1/10 does
not come from a measured stellar mass ratio $M_{1}/M_{2}$, but instead
is inferred from the presence of the afore-mentioned morphological
distortions, which are seen in simulations of mergers of mass ratio
$>$~1/10.

Ideally, one would like to further separate the mergers of type 1 into
major and minor mergers, as defined in $\S$Section~\ref{smdef}.  However,
without individual stellar masses $M_{1}$ and $M_{2}$ for the
progenitors, and with only morphological distortions as a guide, this
is not possible in every case since the morphological disturbances
induced depend not only on the mass ratio of the progenitors, but also
depend to some extent on the orbital geometry (prograde or
retrograde), the gas mass fraction, and structural parameters (e.g.,
Mihos \& Hernquist 1996; Struck 1997; Naab \& Burkert 2001; Mihos et
al. 1995, di Matteo et al. 2007). Therefore, we separate the type 1
mergers into three groups: likely major mergers, likely minor mergers,
and ambiguous cases of `major or minor' mergers as follows:

\begin{description}
\vspace{-3mm} 
\item  
(i) The class of likely major mergers includes systems, which host
fairly unique tell-tale morphological distortions characteristic of a
major merger, such as a train-wreck morphology (e.g., Cases 7, 10, 11,
12 in Figure~\ref{fmer}), or 2 nuclei of similar luminosities. 

\vspace{-3mm} 
\item  
(ii) A system is classified as a likely minor merger if the outer disk
has survived a recent merger.  We base this on the results of
simulations of mergers involving spiral galaxies where the outer disk
of the spiral survives a minor merger, but not a major merger
($\S$Section~\ref{smdef}). This is not true in every case: major mergers of
extremely gas-rich disks with low SF efficiency can lead to a remnant
with an extended stellar disk (Robertson \etal 2004), but such mergers
are unlikely to be relevant for our study, which focuses on
intermediate-mass ($M_{*} \geq 10^{9} M_{\sun}$) systems at redshifts
well below 1.  An additional secondary criterion for classifying a
system as a likely minor merger is that the light from morphological
distortions, tidal debris, or accreted system in the surviving disk is
a small fraction (between 1/4 and 1/10) of the total luminosity of the
disk. Examples include Cases 8, 9, 13 in Figure~\ref{fmer} and
Table~\ref{tmer}.

\vspace{-3mm} 
\item  
(iii) The class of ambiguous `major or minor' merger is assigned to
   systems hosting distortions, which could be due to both a major and
   a minor merger.  Examples are Cases 1--6 in Figure~\ref{fmer} and
   Table~\ref{tmer}.
\end{description}

\vspace{3 mm} {\bf (B) Mergers of type 2a and 2b:} Mergers of type 2a
and 2b are systems, which appear in ACS images as a very close
(separation $d<$ 10 kpc) overlapping pair of two galaxies, and whose
properties suggest they are in the very late phases of a merger of
mass ratio above 1/10.  The difference between mergers of type 2a and
2b is that the latter are resolved into two separate galaxies by the
ground-based COMBO-17 data of resolution $\sim$$1\farcs5$
(corresponding to $\sim$4.3 kpc at $z\sim$~0.165), while the former
are not. Thus, mergers of type 2a only have a single redshift and a
single stellar mass for the pair, while mergers of type 2b have
stellar masses ($M_{1}$, $M_{2}$) and spectrophotometric redshifts
($z_{1}$, $z_{2}$) for both pair members.

We focus only on very close pairs (with $d<$ 10 kpc), and ignore
widely separated pairs for several reasons.  First, in this work, we
are interested in systems with evidence for a {\rm recent} or {\rm
ongoing} merger of mass ratio $>$~1/10.  The very close pairs
represent the very late phases of such a merger, such that
gravitationally bound members can coalesce into a remnant on a short
timescale, comparable to the visibility timescale over which the
morphological distortions in mergers of type 1 persist.  Second, we
expect the morphological distortions from the ongoing merger to be
more prominent in close pairs than widely separated ones.  These
morphological signatures in turn make it easier to separate real pairs
from projection pairs.  Third, as shown by our Monte Carlo
simulations (see below), the statistical contamination from projection
pairs increases significantly for distant pairs.

When identifying potential mergers of type 2b, we consider very close
pairs where one pair member has $M_{*} \geq 10^{9} M_{\sun}$, and its
companion satisfies the following two conditions.  First, the
companion has a mass such that the mass ratio is in the range 1/10 $<
M_{1}/M_{2} \le$ 1/1. Note that this condition allows for pairs where
the companion can be of higher, as well as lower mass (see
Table~\ref{tmer}).  Second, the companion has similar
spectrophotometric redshift, such that the absolute value of ($z_{1} -
z_{2}$) is less than or equal to the 1-sigma redshift error $\sigma_{z}$.
The latter criterion helps to remove some of the false projection
pairs. However, there is still residual contamination by false pairs
as $\sigma_{z}$ is fairly large (see $\S$Section~\ref{sdatas} ; $\sigma_{z}
\sim$~0.028 at $R_{\rm Vega}$~=~23.0).

We identified 5 and 18 potential candidate mergers of type 2a and 2b,
respectively, in our final sample of 886 systems. They are shown in
Figure~\ref{fmer} and Table~\ref{tmer}.  We use the term ``potential
mergers of type 2a and 2b'' when describing the 23 very close ($d<$ 10
kpc) galaxy pairs listed in Table~\ref{tmer} (Cases 14--36), because
these pairs could be a mix of real pairs and false projection pairs
caused by chance line-of-sight superposition. In order to minimize the
contamination from false projection pairs, we only include in our
final analysis those pairs where at least one galaxy shows
morphological distortions indicative of a galaxy-galaxy interaction.
Such pairs are most likely to be real pairs, as opposed to projection
pairs, and we will use these systems to set a firm lower limit to the
merger fraction in $\S$Section~\ref{sinte1}. We find 7 such pairs (Cases
16--18, 20, 24, 34,35), which are marked with a star in column 1 of
Table~\ref{tmer}.  This suggests that the remaining 16/23 or up to
70\% of the types 2a and 2b could potentially be pairs in projection,
due to the dense cluster environment.

 As a second step to gauge the level of contamination from projection
pairs, we performed a Monte Carlo simulation.  We generated a random
galaxy distribution which has an identical redshift distribution to
that of our sample of intermediate-mass ($M_{*} \geq 10^{9} M_{\sun}$)
galaxies, over the STAGES $0.5^{\circ} \times 0.5^{\circ}$ field.  We
ran 100 realizations of this Monte Carlo simulation. For each
realization, we took the ratio ($N_{\rm obs}/N_{\rm mc}$) of observed
to simulated Monte Carlo pairs at different separations ($d$) out to
140 kpc.  The same criteria outlined in $\S$Section~\ref{smvc} to identify
the pairs in the observations are applied to the simulated galaxies.
Namely, we apply the condition that the absolute value of the
difference in spectrophotometric redshifts of the galaxies in a pair
must be less than or equal to the 1-sigma redshift error $\sigma_{z}$,
in order for the galaxies to be counted as a pair.  Figure~\ref{fmc}
shows the ratio ($N_{\rm obs}/N_{\rm mc}$) plotted as a function of
$d$, with the four panels showing the mean, median, minimum, and maximum
values of ($N_{\rm obs}/N_{\rm mc}$). Our results are qualitatively
similar to those found by Kartaltepe \etal (2007).  At large
separations of $d> 50$ kpc, ($N_{\rm obs}/N_{\rm mc}$) is below 1,
and false projection pairs can dominate the statistics.  At small
separations of $d< 20$ kpc, ($N_{\rm obs}/N_{\rm mc}$) is well above
1, suggesting that random chance superposition cannot fully account
for all the observed pairs. Thus, by picking pairs with a separation
of $d< 20$ kpc, one can reduce the contamination from false pairs.
For $d< 20$ kpc, the mean value of the contamination level (defined as
the fraction of observed pairs that are false) ranges from 16\% to
55\% and has a median value of 36\%.  The maximum contamination level
is a factor of $\sim$~1.3 lower than the factor of 70\%, which we
derived in the preceding paragraph, by assuming that all pairs without
morphological distortions are projection pairs. One possible
explanation for the difference is that the afore-described Monte Carlo
simulations do not take into account the large-scale structure and
possible overdensity of galaxies associated with the central parts of
the supercluster.

We next attempt to separate the mergers of type 2a and 2b into major
and minor mergers. For mergers of type 2b, where stellar masses
($M_{1}$, $M_{2}$) are available for each pair member, we can use the
stellar mass ratio $M_{1}/M_{2}$ (see Table~\ref{tmer}).  By
definition, systems with 1/10 $< M_{1}/M_{2} \le$ 1/4 and 1/4 $<
M_{1}/M_{2} \le$ 1 are classified, respectively, as minor and major
mergers.  As shown in Table~\ref{tmer}, potential major mergers of
type 2b are Cases 20-22, 24-26, 28, 29, 32, and 34, while Cases 19,
23, 27, 30, 31, 33, 35, and 36 are potential minor mergers of type 2b.

Mergers of type 2a, however, only have a single redshift and
a single stellar mass, and hence the stellar mass ratio of the pair is
not directly measurable. Instead, we use the stellar light ratio
$L_{1}/L_{2}$, measured from the ACS images, as an approximate proxy
(see Table~\ref{tmer}). Examples of likely major mergers of type 2a  
are Cases 16 and 18 in Figure~\ref{fmer}, while Cases 14, 15, and 17 
are examples of likely minor mergers of type 2a.

\vspace{3mm} {\bf (C) Non-interacting systems of type Symmetric and
Irr-1 :} The remaining systems, which show no indication of a recent
merger, according to the criteria outlined in sections A and B above,
are classified as non-interacting systems. They are subdivided into
two groups called `Non-interacting Irr-1' and  `Non-interacting
Symmetric'.  `Non-interacting Irr-1'
systems exhibit{ \it internally triggered} asymmetries, typically on
scales of a few hundred parsecs.  These asymmetries are generally due
to stochastic SF and/or low ratios of rotational to random velocities.
They can impart a clumpy morphology to a galaxy and are different from
externally-driven distortions, such as tails and ripples, which are
often correlated on scales of several kpc.  The class of
`Non-Interacting Symmetric' consists of galaxies, which are relatively
undistorted. Note that this class includes the undistorted 
galaxies, which are part of these type 2a and 2b pairs identified as 
projection pairs in Section B.   Examples of 
`Non-interacting Irr-1' and ` Non-interacting Symmetric' galaxies 
are shown in Figure~\ref{fexam3}.

In summary after performing the visual classification of the 893
bright ($R_{\rm Vega} \le$~24) intermediate mass ($M_{*} \geq 10^{9}$)
galaxies, we find the following results.  We identify 13 merger
remnants of type 1 (Cases 1--13 in Table~\ref{tmer} and
Figure~\ref{fmer}), and 7 close pairs of type 2a and 2b (Cases 16--18,
20, 24, 34, 35 in Table~\ref{tmer} and Figure~\ref{fmer}) where at
least one galaxy in the close pair shows morphological distortions
indicative of a galaxy-galaxy interaction.  Such pairs are most likely
to be real pairs, as opposed to projection pairs.  We discount those
very close pairs where none of the members are morphologically
distorted.  Our final sample includes 20 mergers and 866
non-interacting galaxies (123 Irr-1 + 743 Symmetric; see
Table~\ref{tvc1}).

\subsection{Tests on the Visual Classes}\label{vctest}

Several tests were performed on the visual classes described in A, B,
and C in $\S$Section~\ref{smvc}.  The visual classifications were calibrated
by having three classifiers (AH, IM, SJ) train on a set of $\sim$100
nearby field galaxies. The full supercluster sample was then
classified by AH and IM, with all ambiguous cases and interacting
galaxies discussed by all three classifiers for final determination.
Random checks on sub-samples of galaxies were performed by SJ.  We
denote as $f_{\rm VC}$ the fraction of intermediate-mass ($M_{*} \geq
10^{9} M_{\sun}$) systems with different visual classes (e.g., mergers
of types 1, 2a, and 2b, and non-interacting systems of type Symmetric
and Irr-1).  In order to gauge the subjectivity inherent in any visual
classification scheme, we estimate the maximum difference in the value
of $f_{\rm VC}$ across the classifiers ($\delta f_{\rm VC}$).  The
maximum percentage difference $P$ defined as ($100 \times \delta
f_{\rm VC}$/$f_{\rm VC}$) is 4\%, 19\%, and 16\% respectively for the
visual classes of Non-interacting Symmetric, Non-interacting Irr-1,
and Mergers.  We conservatively adopt a value of 20\% for $P$ on the
mean merger fraction, as a measure of the inherent subjectivity in the
visual classification.  When citing the final error bar on $f_{\rm
VC}$, we combine in quadrature the term $\delta f_{\rm VC}$, which
represents the subjectivity across classifiers, and the binomial error
term ($\sqrt{f_{\rm VC}(1 - f_{\rm VC})/N}$), where $N$ is the sample
size of 886 systems.  The raw values for $f_{\rm VC}$ are shown in
Table~\ref{tvc1}.

It is also interesting to explore any systematic effects affecting the
mergers we identified visually.  The criteria we used, based on
morphological distortions, might beg the question of whether these
mergers are biased toward a gas-rich population.  The extent to which
a galaxy develops morphological distortions in an interaction depends
on whether the encounter is prograde or retrograde, the interaction
phase, the gas content, and the mass ratio ($M_{1}/M_{2}$).  Di Matteo
\etal (2007) found that the most influential parameters are the mass
ratio and orbital geometry, and not the gas content alone.  The
visibility timescale ($t_{vis}$) on which the morphological signatures
persist also depend on the orbital geometry, gas content, and the mass
ratio.  We do not have a direct measurement of the gas content of
galaxies in A901/902, so we attempt an indirect first-order test in
order to verify whether our methodology preferentially picks gas-rich
mergers.  We visually classify sample galaxies into three broad classes:
`highly smooth', `highly clumpy', and `other'.  If a system shows a
high degree of clumpiness (i.e. very patchy regions that are likely
associated with significant amounts of gas, dust, and SF), it is
classified as `highly clumpy'. Conversely, systems with a very smooth
non-clumpy appearance are classified as `highly smooth'.

Out of the 20 mergers and 866 non-interacting galaxies, we find that
6/20 (30\%) and 210/866 (24\%) are classified as `highly clumpy',
respectively.  The comparable fraction of `highly clumpy' systems
among non-interacting systems and mergers suggests that our method for
identifying mergers does not show a strong bias toward preferentially
selecting mergers, which are highly clumpy, gas-rich, and
star-forming.  It is also interesting to recall here that our visual
classification scheme takes special care to distinguish between
Non-interacting Irr-1 and mergers ($\S$Section~\ref{smvc}). The
Non-interacting Irr-1 galaxies were defined as those where
internally triggered small-scale asymmetries, often due to SF, are
present. Thus, as expected, if we separate our non-interacting
galaxies into Irr-1 and Symmetric, we find that a large fraction of
the Non-interacting Irr-1 (111/123 or 90\%) are `highly clumpy',
compared to 99/713 (14\%) of the Non-interacting Symmetric systems.

\subsection{CAS: Quantitative method for capturing interacting galaxies}\label{smcas1}

Using the CAS code (Conselice \etal 2000, Conselice 2003a),
quantitative structural parameters measuring the concentration ($C$),
asymmetry ($A$), and clumpiness ($S$) were derived for the supercluster
galaxies.  CAS was run on the F606W images and the segmentation maps
produced during the original source extraction (Caldwell \etal 2008)
were used to mask neighboring galaxies.

The CAS  concentration  index $C$ (Bershady \etal  2000) is proportional 
to  the logarithm of the ratio of the 80\%--20\% curve of growth radii 
within 1.5 times the Petrosian inverted radius at $r$($\eta$~=~0.2):
\begin{equation}
C = 5 \times {\rm log}\left(\!\frac{r_{80\%}}{r_{20\%}}\!\right).
\end{equation}

The CAS code derives the asymmetry index $A$ (Conselice 2003) by
rotating a galaxy image by 180 deg, subtracting the rotated image from
the original image, summing the absolute intensities of the residuals,
and normalizing the sum to the original galaxy flux.  CAS improves the
initial input center with the IRAF task `imcenter' and then performs a
further refinement within a $3 \times~3$ grid, picking the center that
minimizes $A$.

The clumpiness parameter $S$ is a quantitative measure of the high
spatial frequency patchiness of a galaxy.  It is defined as the
summation of the difference of the galaxy's original flux and the flux
of an image with the high frequency structures muted by smoothing over
a filter on order of the clumpiness, taking into account the
background flux over that same area.  This measurement is then divided
by the summation of the original flux of the galaxy to obtain the
clumpiness parameter $S$.

Conselice \etal (2000) argue that the CAS merger criterion ($A>S$ and
$A>$~0.35) at $\lambda_{\rm rest} > $~4500 \AA \, can be used to
capture galaxies which have strong asymmetries indicative of
major mergers.  However, calibrations of the asymmetry value with
N-body simulations (Conselice 2006; Lotz \etal 2008) have shown that
for galaxies involved in major mergers of mass ratios 1:1--1:3, the
asymmetries vary as a function of time during the merger.  The $A$
value reaches a peak near the midpoint of the merger and falls off
both before and after this time to less than 0.35.  The criterion
$A$~$>$~0.35 is fulfilled only $\sim$1/3 of the time for major mergers in
these simulations, while minor mergers of mass ratios below 1:5 have
$A$ values significantly lower than 0.35.

Furthermore, in intermediate mass ($M_{*} \geq 10^{9} M_{\sun}$) field
galaxies at $z\sim$~0.24--0.80, recent work (Jogee \etal 2008, 2009;
Miller \etal 2008) has shown that the CAS criterion typically only
picks up 50\%--70$\%$ of galaxies visually typed as being disturbed
or interacting. These results caution against using CAS without
complementary visual classifications in field galaxies.

The effectiveness of the CAS criterion has not been explored in
cluster environments where the dominant galaxy population is more
gas-poor than in the field.  In $\S$Section~\ref{scas1}, we present the
merger fraction ($f_{\rm CAS}$) from CAS and perform one of the first
systematic comparisons to date between CAS-based and visual
classification results in clusters.

\section{Results and Discussion}\label{sresul}

\subsection{Merger fraction in A901/902 from visual classification}\label{sinte1}

Next we discuss how to define and estimate the merger fraction.  Our
goal is to estimate the fraction of $f_{\rm merge}$ of systems with
stellar mass above an appropriately chosen mass cut $M_{\rm cut}$,
which are involved in mergers of mass ratio $\ge$~1/10.  The merger
fraction $f_{\rm merge}$ is computed as ($N_{\rm merge}/N_{\rm tot}$),
where $N_{\rm merge}$ is the number of major and minor mergers
involving galaxies with $M_{*} \ge M_{\rm cut}$, and $N_{\rm tot}$ is
the total number of galaxies with $M_{*} \ge M_{\rm cut}$.

It is important to determine the minimum stellar masses of the major
and minor mergers involving galaxies with $M_{*} \ge M_{\rm cut}$ and
thereby assess for what value of $M_{\rm cut}$ we can trace such
mergers.  Major merger pairs (defined as having mass ratio 1/4 $<
M_{1}/M_{2} \le$ 1) of mass ratio 1:1 to 1:3 will have minimum stellar
masses ranging from $2 \times M_{\rm cut}$ to $4 \times M_{\rm cut}$,
in cases where galaxies of mass $M_{*} \ge M_{\rm cut}$ merge with
systems at least as massive as themselves.  However, if galaxies of
mass $M_{*} \ge M_{\rm cut}$ merge with lower mass systems, the
minimum mass of 1:1 to 1:3 mergers will range from $2 \times M_{\rm
cut}$ to ($4 \times M_{\rm cut}/3$).  Taken together, the above
constraints imply that we are complete for major mergers involving
galaxies with $M_{*} \ge M_{\rm cut}$, as long as we can trace {\rm
1:3 major mergers with a minimum stellar mass of ($4 \times M_{\rm
cut}/3$).

Repeating the exercise for minor mergers (defined as having 1/10 $<
M_{1}/M_{2} \le$ 1/4), it follows that the minimum stellar mass for
1:4 to 1:9 mergers ranges from $5 \times M_{\rm cut}$ to $9 \times
M_{\rm cut}$ or from ($5 \times M_{\rm cut}/4$) to ($10 \times M_{\rm
cut}/9$), depending on whether galaxies of mass $M_{*} \ge M_{\rm
cut}$ merge with systems of higher mass or lower mass.  Thus, we are
complete for minor mergers involving galaxies with $M_{*} \ge M_{\rm
cut}$, as long as we can trace {\rm 1:9 minor mergers with a minimum
stellar mass of ($10 \times M_{\rm cut}/9$).

For what value of $M_{\rm cut}$ are these conditions satisfied by the
mergers of Types 1, 2a, and 2b, which we visually identified in the
final sample ($\S$Section~\ref{smvc})?  Since our final sample is complete
for $M_{*} \ge 10^{9} M_{\sun}$, it follows that we can identify all
mergers of type 1 and 2a with $M_{*} \ge 10^{9} M_{\sun}$ from this
sample.  If we impose this value to the afore-defined criteria of ($10
\times M_{\rm cut}/9$) and ($4 \times M_{\rm cut}/3$), it implies that
a mass cut $ M_{\rm cut} \ge 0.9 \times 10^{9} M_{\sun}$ would allow
us to be complete for major and minor mergers of type 1 and 2a.
However, the situation is different for the mergers of type 2b (close
pairs resolved into two galaxies by COMBO-17) because the individual
galaxies making up the pair are complete only for $M _{*} \geq10^{9}$
$M_{\sun}$. As a result, we can only completely trace 1:3 major
mergers of type 2b with total mass $\ge 4.0 \times 10^{9} M_{\sun}$.
If we impose this value to the afore-defined criterion of ($4 \times
M_{\rm cut}/3$)}, it implies that a mass cut of $ M_{\rm cut} \ge 3.0
\times 10^{9} M_{\sun}$ is needed to ensure completeness for major
mergers of type 2b.  Similarly, we can only completely trace 1:9 minor
mergers of type 2b with total mass $ \ge 9.0 \times 10^{9} M_{\sun}$,
which implies that a mass cut $M_{\rm cut} \ge 9.0 \times 10^{9} M_{\sun}$
is needed to completely trace all minor mergers of type 2b.

Using a mass cut $M_{\rm cut} \ge 9.0 \times 10^{9} M_{\sun}$ for
computing the merger fraction $f_{\rm merge}$ would allow us to be
complete for major and minor mergers of types 1, 2a, and 2b.  However,
it leads to very small number statistics and is not viable.  We
therefore explore the value of $f_{\rm merge}$ for the less severe cut
of $M_{\rm cut} \ge 3.0 \times 10^{9} M_{\sun}$, as well as for
$M_{\rm cut} \ge 10^{9} M_{\sun}$.  The cut of $M_{\rm cut} \ge 3.0
\times 10^{9} M_{\sun}$ ensures that we are complete for major and
minor mergers of type 1, 2a, but leaves us incomplete for mergers of
type 2b.  It is encouraging that both cuts yield similar values for
the merger fraction $f_{\rm merge}$, as illustrated in
Table~\ref{tcut}: the merger fraction is 0.023$\pm$0.007 and
0.021$\pm$0.007, respectively, for $M_{\rm cut} \ge 10^{9} M_{\sun}$
and $M_{\rm cut} \ge 3.0 \times 10^{9} M_{\sun}$.  Note that in
computing the merger fraction, we only include the 20 distorted
mergers listed in Table~\ref{tmer}, and avoid the potential projection
pairs of type 2a and 2b without signs of morphological distortions
\footnote{Even if we included all pairs of type 2a and 2b, the merger
fraction would still have similar values (0.041$\pm$0.01 and
0.033$\pm$0.01) for both mass cuts (Table~\ref{tcut}).}.

While mergers may have played an important role in the evolution of
cluster galaxies at earlier times (e.g., $z>2$), hierarchical models
(e.g., Gottloeber et al 2001; Khochfar \& Burkert 2001) predict that
the merger fraction in dense clusters falls more steeply at $z<1$ than
the field merger fraction.  As a result, at $z<0.3$, the merger
fraction for intermediate mass cluster galaxies is predicted to be
quite low (typically below 5\%).  The low merger fraction among
intermediate mass ($M = 10^{9}$ to a few $\times 10^{10} M_{\sun}$)
galaxies in the A901/902 clusters is consistent with the latter
prediction.

How are the mergers distributed among major mergers, minor mergers,
and ambiguous cases that could be either major or minor mergers?  The
results are shown in Columns 8--10 of Table ~\ref{tcut}, based on
the classification listed in Column 8 of Table ~\ref{tmer}.  The
estimated fractions of likely major mergers, likely minor mergers, and
ambiguous cases are 0.01$\pm$0.004\% (9/886), 0.006$\pm$0.003\%
(5/886), and 0.007$\pm$0.003\% (6/886), respectively for $M_{\rm cut}
\ge 10^{9} M_{\sun}$.  For $M_{\rm cut} \ge 3.0 \times 10^{9}
M_{\sun}$, the corresponding fractions are 0.013$\pm$ 0.005\% (8/609),
0.008$\pm$0.004\% (5/609), and 0.0\% (0/609), respectively.

In the rest of this paper, we continue to work with a mass cut of
$M_{\rm cut} \ge 10^{9} M_{\sun}$, and the 20 distorted mergers
(Table~\ref{tmer}) applicable for this mass cut.  However, where
relevant, we cite many of our results for both of the mass cuts
($M_{\rm cut} \ge 10^{9} M_{\sun}$ and $M_{\rm cut} \ge 3.0 \times
10^{9} M_{\sun}$) so that we can gauge the potential effect of
incompleteness in tracing major and minor mergers.

\subsection{Frequency of mergers in A901/902 from CAS}\label{scas1}
The results of running CAS on the final classified sample of 886
 intermediate-mass ($M_{*} \geq 10^{9}$ $M_{\sun}$) systems are shown
 in Table~\ref{tcas} and Figures~\ref{fcas1} and~\ref{fcas2}.  In
 Figure~\ref{fcas1}, the 20 mergers of type 1, 2a, and 2b are plotted
 in different symbols.  For Type 2b merger pairs, we plot the highest
 value of CAS $A$ found between the two galaxies in each system.  Using
 the CAS merger criterion ($A>0.35$ and $A>S$) to identify mergers
 yields a merger fraction ($f_{\rm CAS}$) in this sample of 18/886 or
 $0.02\pm 0.006$.  For the more conservative sample with a mass cut
 $M_{\rm cut} \ge 3 \times 10^{9} M_{\sun}$, $f_{\rm CAS}$ is 7/609 or
 $0.011\pm0.005$.

When citing the error on $f_{\rm CAS}$, we take the largest of either
the Poisson error or the systematic error ($\sigma_{\rm CAS}$) in
$f_{\rm CAS}$ due to the systematic errors in CAS $A$ and $S$.  The
systematic error, $\sigma_{\rm CAS}$, is calculated by taking the
upper and lower bounds of $f_{\rm CAS}$ based on ($ A\pm$ error in $A$)
and ($S \pm$ error in $S$). Specifically, these limits on $f_{\rm CAS}$
are found by using the criteria of (($A \pm$ error in $A$) $<$ ($S \pm$
error in $S$)) and (($A \pm$ error in $A$) $>$ 0.35).

At first sight, the CAS-based merger fractions $f_{\rm CAS}$ for the
two mass cuts ($0.020\pm0.006$ and $0.011\pm0.005$ for $M_{\rm cut}
\ge 10^{9} M_{\sun}$ and $M_{\rm cut} \ge 3\times 10^{9} M_{\sun}$,
respectively) are not widely different from the merger fraction
$f_{\rm merge}$ based on visual classification (0.023$\pm$0.007 and
0.021$\pm$0.007 for $M_{\rm cut} \ge 10^{9} M_{\sun}$ and $M_{\rm cut}
\ge 3\times 10^{9} M_{\sun}$, respectively; $\S$Section~\ref{sinte1}).
However, the comparison of $f_{\rm CAS}$ and $f_{\rm merge}$ does not
tell the whole story because the nature of the systems picked by the
two methods can be quite different.  The visual classes of the 18
systems, which satisfy the CAS criterion and are considered as mergers
in the CAS system, are shown in Table~\ref{tcas}.  It turns out 7/18
($39\pm14\%$) of these ``CAS mergers'' are visually classified as
non-interacting systems.  The results are illustrated in
Table~\ref{tcas} and Figure~\ref{fcas1}.  Figure~\ref{fcas2} shows
examples of these ``contaminants'': they tend to be dusty highly
inclined systems and systems with low level asymmetries that seem to
be caused by SF.

It is also useful to ask what fraction of the 20 systems
visually-identified as mergers satisfy the CAS criterion.  We find
that for $M_{\rm cut} \ge 10^{9} M_{\sun}$, the CAS criterion only
captures 11 of the 20 ($55\pm16\%$) of the visually-classified
mergers. These results are illustrated in Figure~\ref{fcas1}.
Figure~\ref{fcas2} shows examples of merging galaxies missed by CAS.
The missed mergers include galaxies with fainter outer tidal features,
double nuclei where CAS puts the center between the nuclei, and pairs
of very close connected galaxies.

It is also interesting to ask what percentage of the different Types
of mergers (Types 1, 2a, and 2b) does the CAS merger criterion
recover.  Figure~\ref{fcas1} also shows the mergers divided up by
merger types 1, 2a and 2b.  Of the 13 mergers of type 1, CAS recovers
9/13 or 69\%$\pm$18\%. Of the 3 mergers of type 2a, CAS recovers 1/3 or
33\%$\pm$28\%. Of the 4 mergers of type 2b, CAS recovers 1/4 or
25\%$\pm$22\%.

Since the CAS criterion is widely used to pick major mergers, it is
also interesting to explore how well this criterion picks up the
systems that we visually classified as major mergers
(Table~\ref{tmer}).  We find that the CAS merger criterion picks up
6/9 (67\%$\pm20\%$) of the systems classified as major mergers.  It is
also interesting to note that the CAS criterion recovers 2/5
(40\%$\pm23\%$) and 4/6 (67\%$\pm23\%$) of the systems classified
respectively as minor mergers and ambiguous mergers.

\subsection{Distribution of  mergers}\label{sinte2}

In order to define different regions of the A901a, A901b, and A902
clusters, we computed the projected number density $n$
(Figure~\ref{fnumd}) for intermediate-mass ($M_{*} \geq 10^{9}
M_{\sun}$) galaxies as a function of clustocentric radius $R$ by
assuming a spherical distribution.  Note here that each galaxy is
assigned to the cluster closest to it, and $R$ is measured from the
center of that cluster.

We consider the cluster core to be at $R\le$~0.25 Mpc, as this is the
region where the projected number density $n$ rises very steeply
(Figure~\ref{fnumd}).  The cluster virial radii are taken to be
$\sim$1.2 Mpc, based on estimates from the DM maps derived
from gravitational lensing by Heymans \etal (2008).  Throughout this
paper, we refer to the region at 0.25 Mpc~$<R \le$~1.2 Mpc, between
the cluster core and the cluster virial radius, as the `outer region
of the cluster'. The region outside the virial radius (1.2 Mpc~$< R
\le$~2.0 Mpc) is referred to as the `outskirt region of the
cluster'. The core, outer region, and outskirt region are labeled on
Figure~\ref{fnumd}.

As shown in Table~\ref{tcompa}, the A901 clusters have central galaxy
number densities ($n$) in the core region of 1000--1600 galaxies
Mpc$^{-3}$. These are lower than that of the rich Coma cluster and
higher than that of the Virgo cluster.

Figure~\ref{fdist1} shows the spatial distribution of the mergers
visually identified among the sample of intermediate-mass ($M_{*} \geq
10^{9} M_{\sun}$) galaxies in the A901a, A901b, and A902 clusters.  We
again consider here only the 20 distorted mergers listed in
Table~\ref{tmer}, for the reasons discussed in $\S$Section~\ref{sinte1}.  We
find that all the 20 mergers lie outside the cluster cores.  We
estimate that the number density ($n_{\rm merge}$) of mergers is 0,
2.09, and 0.11 galaxies Mpc$^{-3}$, respectively, in the core, outer
region, and outskirt of the cluster (Table~\ref{t3rad}).  When
estimating $n_{\rm merge}$ in the three regions, we divided the number
of mergers by the volume of the core region, and the volume of a
spherical shell in the outer region and outskirt of the clusters.

For mergers and non-interacting galaxies, we compute the minimum
distance to the nearest cluster center (A901a, A901b, and A902), and
the local values of various environmental parameters, such as the
local galaxy surface density $\Sigma_{\rm 10}$, the local DM mass surface
density $\kappa$, and the relative local ICM density.  Following Wolf
\etal (2005) and Gilmour \etal (2007), we define and compute the local
galaxy surface density ($\Sigma_{\rm 10}$) as the number of galaxies
per (Mpc/h)$^{-2}$ that are in an annulus bracketed by two circles whose
radii are equal to the average distances to the 9th and 10th nearest
neighbor.  The DM mass surface density ($\kappa$) is computed locally
using the coordinates of each galaxy on the weak lensing map provided
by Heymans \etal (2008). The relative local ICM counts are measured at
each galaxy position using the X-ray map from Gray \etal (2010, in preparation) and
Gilmour \etal (2007).  Figure~\ref{fnsigk} shows these local
environmental parameters as a function of the minimum distance to the
cluster center for mergers and non-interacting galaxies.  Since the
mergers lie outside the cluster core (0.25 Mpc~$< R \le$~2 Mpc), they
are associated with low values of $\kappa$ and intermediate values of
$\Sigma_{\rm 10}$, and ICM density (Figure~\ref{fnsigk}).

\subsection{What accounts for the distribution of mergers?}\label{sinte3}

The timescale for collisions and close encounters is given by:
\begin{equation}
t_{\rm coll} = \frac{1}{n \sigma_{\rm gal} A},
\end{equation}

where $n$ is the galaxy number density, $\sigma_{\rm gal}$ is the
galaxy velocity dispersion, $A$ is the collisional cross section ($\pi
f(2r_{\rm gal})^{2}$), $r_{gal}$ is a typical galaxy radius, and $f$ is
the gravitational focusing factor (Binney \& Tremaine, 1987). Since
the average velocity dispersion in clusters is high, the value for $f$
will be low and we will assume it to be on the order of unity. Thus,

\begin{equation}\label{tcoll}
\begin{split}
t_{\rm coll}\sim800\! \times
\left(\!\frac{n}{10^{3} \rm \, {\rm  Mpc}^{-3}}\!\right)^{-1}\! \times
\left(\frac{\sigma}{10^{3}\,{\rm km\,s}^{-1}}\right)^{-1}\! \\ \times
\left(\!\frac{{r_{\rm gal}}}{10 \, {{\rm kpc}}}\!\right )^{-2}\!
{\rm Myr}
\end{split}
\end{equation}

We compute $n$ using the same method as in $\S$Section~\ref{sinte2}
for the core, and use spherical shells to compute $n$ in the outer
region and cluster outskirt for our sample of $R_{\rm Vega} \le$~24,
intermediate mass ($M_{*} \geq10^{9} M_{\sun}$) galaxies. The local
galaxy velocity dispersion profiles for A901a/b, A902, and the South
West Group (SWG) from kinematic modeling using $\sim$420 2dF redshifts
(Gray \etal 2010, in preparation) are shown in Figure~\ref{fcvdp}, and the average
velocity dispersions $\sigma_{\rm gal}$ are shown in
Table~\ref{tcompa}.

The central galaxy velocity dispersion within the cores ($R<0.25$~Mpc)
of A901a,b and A902 typically range from 700 to 1000 km s$^{-1}$.
Beyond the cluster core, in the outer region (0.25 Mpc~$< R \le$~1.2
Mpc), the small number statistics leads to large error bars on the
galaxy velocity dispersion, making it not viable to determine whether
it remains high, drops, or rises. To estimate timescales, we take the
average $\sigma_{\rm gal}$ for A901/902 to be $\sim$800 km s$^{-1}$ in
three regions of the cluster.  The timescales ($t_{\rm coll}$) for close
encounters are shown in Table~\ref{t3rad} and range from 0.765 to 13.5
to 135 Gyrs from the core ($R\le$~0.25 Mpc), to the outer region (0.25
Mpc~$< R \le$~1.2 Mpc), to the outskirt region (1.2 Mpc~$< R \le$~2.0
Mpc) of the cluster.

It may at first seem surprising that mergers do not preferentially lie
in the cluster core, where the probability for close encounters is
high due to the associated short value of the timescale $t_{\rm coll}$ for
collisions and close encounters. However, if $\sigma_{\rm gal}$ is large,
a close encounter is unlikely to lead to a merger or a large amount of
tidal damage (i.e. to strong morphological and kinematical
distortions).  This is because a large galaxy velocity dispersion will
likely cause the energy {\it E} of the reduced particle in a binary
encounter to be positive, causing the orbit to become unbound (Binney
\& Tremaine, 1987).

All merging systems lie at a projected clustocentric radius of 0.25
Mpc~$<R \le$~2 Mpc, between the core and cluster outskirts, although in
this region the timescale for collisions and close encounters is quite
large.  We suggest two possible reasons why strong interactions and
mergers might populate the outer region and cluster outskirt. The
first possibility is that the velocity dispersion of galaxies falls
between the core and outskirt, due to the {\it intrinsic} velocity
dispersion profile of the clusters.  Many clusters show a flat
velocity dispersion profile from 1 Mpc outward, but several also show
a declining profile (den Hartog \& Katgert 1996, Carlberg 1997). In
the latter cases, the velocity dispersions can fall from 1500 to 100
km s$^{-1}$ from the cluster core to cluster outskirt.  Since the
local velocity dispersion profiles in the regions of interest in
A901/902 (Figure~\ref{fcvdp}) are based on a small number of galaxies
and therefore large errors in $\sigma_{\rm gal}$, we cannot determine if
the velocity dispersion profile falls or remains flat.  An additional
difficulty is that the member galaxies of the neighboring clusters
influence $\sigma_{\rm gal}$ at $\sim$0.5 Mpc for A901a/b and $\sim$1
Mpc for A902.

The second possibility is that the mergers are part of groups or field
galaxies that are being accreted along filaments by the A901/902
clusters.  This would be in line with the idea that clusters
continuously grow by mergers and accretion of groups (e.g., Zabludoff
\& Franx 1993; Abraham \etal 1996a; Balogh, Navarro \& Morris 2000).
Groups have lower galaxy velocity dispersion (typically below 300 km
s$^{-1}$) than clusters, as shown in Table~\ref{t3rad}.  Furthermore
simulations show that merger rates are highest as field and group
galaxies accrete into cluster along cosmological filaments (see below;
van Kampen \& Katgert 1997; Figure~\ref{felco1}).

To further quantify the possibility of groups accreting into the
A901/A902 clusters, we compare the data to model predictions from
simulations of the STAGES A901/A902 supercluster (van Kampen \etal in
preparation).  The simulation aims to reproduce the environment as observed
for A901/A902 by using constrained initial conditions that produce
clusters with overall properties similar to those of A901, A902, and
the neighboring A907 and A868.  The simulation box is large enough to
contain all these, and hence reproduce the overall large-scale density
and velocity fields similar to what is observed. A phenomenological
galaxy formation model is then used to simulate the galaxy population
in the same box, allowing us to select mock galaxies above a certain
luminosity or stellar mass cut, as done in the observed dataset.  A
stellar mass cut of $M_{*}\geq$ 10$^{9}$ $M_{\sun}$ is applied in the
simulations, corresponding to the mass cut in the observational
sample. The solid lines in Figure~\ref{felco1} show the predicted
number density ($n$) and fraction ($f$) of galaxy mergers with mass
ratio $M_{1}/M_{2} >$~1/10, as a function of environment, as
characterized by the local overdensity ($\delta^{\rm G}$).  The latter
is calculated by smoothing the density of DM halos with a
Gaussian of width 0.4 Mpc to take out the effect of individual
galaxies.  Typical values of $\delta^{\rm G}$ are $\sim$10-100 for
group overdensities, $\sim$200 at the cluster virial radius, and
$\gtrsim$~1000 in the core of rich clusters.

Typically, in the simulations, as field and group galaxies fall into a
cluster along filaments, the coherent bulk flow enhances the galaxy
density and causes galaxies to have small relative velocities, thus
leading to a high probability for mergers at typical group
overdensities.  Closer to the cluster core, galaxies show large random
motions rather than bulk flow motion, leading to a large galaxy
velocity dispersion, and a sharp drop in the probability of mergers.

In order to compare the data to simulation results, we use the number
($N_{\rm merge}$) and fraction ($f_{\rm merge}$) of mergers of mass
ratio $\ge 1/10$, which we computed in $\S$Section~\ref{sinte1}.  The solid
curve in Figure~\ref{felco1} shows the merger fraction and number
density predicted by the simulations as a function of overdensity.
The three dashed lines in Figure~\ref{felco1} show the estimated range in
the observational number density ($n_{\rm merge}$) and fraction
($f_{\rm merge}$) of mergers in the three different regions (the core, the
outer region, and the outskirt) of the A901/902 clusters, as defined
in $\S$Section~\ref{sinte2}.  The points at which the dashed lines cross or
approach the solid curve tell us the typical overdensities at which we
expect to find such merger fraction or merger number densities in the
simulations.  It can be seen that the low merger density seen in the
core region of the cluster correspond to those expected at typical
cluster core overdensities.  On the other hand, the larger merger
fraction we observe between the cluster core and outer region (0.25
Mpc~$< R \le$~2 Mpc) is close to those seen in typical {\it group
overdensities}.  Our results are in agreement with the above scenario,
whereby the accretion of group and/or field galaxies along filaments,
where they have low relative velocities and enhanced overdensities,
leads to enhanced merger rates.  We note that similar conclusions are
reached if we perform the comparisons between data and simulations
using the more conservative mass cut of $M_{*} \geq 3 \times 10^{9}
M_{\sun}$, discussed in $\S$Section~\ref{sinte1}.

However, there are a couple of caveats when comparing our data with
simulations. One caveat is that we directly compare the projected
values of the number density or fraction of mergers from two
dimensional observations, to the predicted number density of mergers
from three dimensional simulations.  Projection effects can introduce
uncertainties in our observational estimate. The magnitude of the
uncertainties due to projection effects depends on the detailed
environment and will be investigated in the full STAGES supercluster
simulation dataset (van Kampen \etal 2009, in preparation).

A further indirect evidence for group accretion stems from  
comparing semi-analytic galaxy catalogs to STAGES observations. 
Rhodes \etal (in preparation) finds an overabundance in galaxies in A902 
compared to its mass. This could be explained by  two or more galaxy 
groups in projection, consistent with the idea of group accretion.

\subsection{Comparison with groups and clusters at different epochs}\label{scomp1}

We first recapitulate our results on the visually-based merger
fraction ($f_{\rm merge}$) in the A901/902 clusters
($\S$Section~\ref{sinte1}).  Among intermediate-mass ($M_{*} \geq10^{9}
M_{\sun}$) systems, we find that the fraction $f_{\rm merge}$ of
systems which show evidence of a recent or ongoing merger of mass
ratio $ >$~1/10 is 0.023$\pm$0.007 (Table~\ref{tcut}).  Most of these
mergers have $M_{\rm V} \sim -19$ to $-22$ and $L \le L^{*}$ (see
Figure~\ref{flumf}).  We also estimated that the fraction of likely
major mergers, likely minor mergers, and ambiguous cases to be
0.01$\pm$0.004\% (9/886), 0.006$\pm$0.003\% (5/886), and
0.007$\pm$0.003\% (6/886), respectively.

Next, we compare our merger fraction in the A901/902 clusters with the
published merger fraction in other clusters and group galaxies out to
$z\sim$~0.8, over the last 7 Gyr (Figure~\ref{fcompa}). These
comparisons are difficult to make as different studies apply different
luminosity or mass cuts.  Furthermore, some studies consider only
major mergers, while others consider all interacting galaxies, which
are likely candidates for both major and minor interactions.

The variation in galaxy populations sampled at different epochs must
also be kept in mind when comparing results at lower redshift with
those out to $z\sim$~0.8.  Low redshift samples typically sample a
small volume and therefore tend to host only a small number of the
brightest and most massive systems.  Conversely, higher redshift
magnitude-limited samples will suffer from Malmquist bias and tend to
under-represent the faint ($L < L^{*}$) galaxies.

There have been several studies of galaxy mergers and interactions in
intermediate redshift clusters (Lavery \& Henry 1988; Lavery, Pierce,
\& Mclure 1992; Dressler \etal 1994; Oemler, Dressler, \& Butcher
1997; Couch \etal 1998), but only few to date with high resolution
$HST$ imaging.  Couch \etal (1998) used WFPC/WFPC2 observations and
spectroscopy of two clusters at $z\sim$0.3 and $\sim0.4$, focusing on
galaxies with $R\sim$22.5 or $L < L^{*}$.  In their study, they
classified $\sim$7/200 or $\sim3.5\%$ galaxies to be merging based on
separations and visible distortions. The merger fractions of these two
intermediate redshift clusters are plotted in Figure~\ref{fcompa}.

It is also interesting to note that the merging systems in these two
intermediate redshift clusters tend to be blue and preferentially
located in the outskirt of the clusters.  In fact, in the Couch \etal
(1998) sample, $\sim60\%$ of the merging galaxies among $L <L^{*}$
systems are blue.  The mergers in these two intermediate redshift
clusters are similar in many ways to the mergers in the A901/902
clusters: the latter mergers lie in between the cluster core and
outskirt (0.25 Mpc~$< R \le$~2 Mpc; $\S$Section~\ref{sinte2}) of the A901/902
clusters, and for the absolute magnitude range of $M_{\rm V} \sim$-19
to -22, 16/20 or $80 \pm 18\%$ of these mergers lie on the blue cloud
($\S$Section~\ref{scolor1}).

In a study by Dressler \etal (1994), based on WFPC images of a
$z\sim0.4$ cluster, the fraction of bright ($M_{V} < -18.5$) galaxies
identified as mergers was 10/135.  Another WFPC study of four clusters
at $z=0.37-0.41$ by Oemler, Dressler, \& Butcher (1997) placed a lower
limit of 5\% on the fraction of merging galaxies with $M_{V} < -19$ or
$L \leq L^{*}$.

A high redshift ($z=0.83$) cluster was studied by van Dokkum \etal
(1999) using $HST$ WFPC2 images and spectroscopic redshifts for 80
confirmed members.  They found a high fraction (17\%) of luminous
($M_{B}\sim -22$; $L \sim 2L^{*}$) systems, which exhibit signatures
of ongoing mergers.  These mergers were found to be red,
bulge-dominated, and located in the outskirts of the cluster.  A later
study by Tran \etal (2005b) used spectroscopic followup to confirm
that 15.7\%$\pm3.6\%$ of these galaxies were bound pairs with
separations $\ltsim$30h$^{-1}$ kpc and relative velocities $\ltsim$300
km s$^{-1}$. These merging pairs are located outside the cluster
centers, similar to the mergers in A901/902. However, the galaxy
sample at high redshift is dominated by more luminous galaxies
($M_{B}\sim$ -22; $L \sim2L^{*}$) than the A901/902 sample where most
galaxies have $M_{V}\sim$ $-$19.5 to $-$22 (Figure~\ref{flumf}).  The
number statistics for bright ($\sim 2L^{*}$) mergers in A901/902 are
not robust enough to allow a direct comparison with the merger
fraction cited in van Dokkum \etal (1999).

Finally, we look at the merger fraction in {\it groups}.  A study by
McIntosh \etal (2008) finds that $\sim$1.5\% of massive ($M \ge 5
\times 10^{10} M_{\sun}$) galaxies in SDSS groups (with $M_{\rm halo}
> 2.5 \times 10^{13}$ $M_{\sun}$) are major mergers at $z$ of
0.01--0.12.  Most of the mergers involve two red sequence galaxies and
they are located between 0.2 and 0.5 Mpc from the group center.  A
study by McGee \etal (2008) also finds an enhancement in asymmetric
bulge-dominated galaxies in groups, consistent with a higher
probability for merging in the group environment.  Additional evidence
for mergers in groups is shown in a study of a supergroup at
$z\sim0.37$ by Tran \etal (2008), who report dry dissipationless
mergers and signatures thereof in three of four brightest group
galaxies.

A merger fraction of 6\% was found by Zepf (1993) in Hickson compact
groups at $z< 0.05$, among systems with luminosities $L \leq L^{*}$.
The merger fraction is significantly higher than that in SDSS
groups. The increased merger fraction is expected since Hickson
compact groups are different from loose groups: they have high number
densities comparable to those in cluster cores, but low galaxy
velocity dispersions. These two conditions provide an environment most
favorable to strong tidal interactions and mergers, as argued in
$\S$Section~\ref{sinte3}.

It is not straightforward to draw conclusions about the evolution of
the merger fraction in cluster galaxies over $z\sim$~0.05--1.0 from
the above studies because they sample different types of systems and
different luminosity ranges. If we conservatively restrict ourselves
to only studies of $L \leq L^{*}$ cluster galaxies, then we have 4
data points over $z\sim$~0.17--0.4, shown as solid filled circles in
Figure~\ref{fcompa}.  The mean value of the data points allows for
evolution by a factor of $\sim$3.2, in the merger fraction of $L \leq
L^{*}$ cluster galaxies over $z\sim$~0.17--0.4.  However, if one
uses the full range of merger fractions allowed by the error bars on
the data points, then we can admit a wider spectrum of scenarios,
ranging from no evolution to evolution by a factor of $\sim$5 over
$z\sim$~0.17--0.4.  Having additional deep, large-volume, high
resolution studies, which are based on larger samples with smaller
error bars would help to separate between these scenarios, and thereby
test hierarchical models of galaxy evolution.

As mentioned in $\S$Section~\ref{sinte1}, hierarchical models (e.g.,
Gottloeber et al 2001; Khochfar \& Burkert 2001) predict that the
merger fraction in dense clusters falls more steeply at $z<1$ than the
field merger fraction, such that at $z<0.3$, the predicted merger
fraction is quite low (typically below 5\%) among intermediate mass
cluster galaxies.  The low merger fractions among intermediate mass
($10^{9}$ to a few $\times 10^{10} M_{\sun}$) or intermediate
luminosities ($L < L^{*}$) galaxies in the A901/902 clusters and other
low redshift clusters are consistent with these predictions, but we
cannot yet test the predicted rate of evolution of the merger fraction
in clusters with redshfit.

\subsection{Galaxies on the blue cloud and red sample}\label{scolor1}

We explore the properties of galaxies on the blue cloud and in the red
sample among the final sample of 886 systems (20 mergers and 866 
non-interacting galaxies; $\S$Section~\ref{smvc}). The red sample was defined
in Wolf \etal (2009) and contains both passively evolving spheroidal
galaxies on the red sequence, as well as dusty red galaxies that lie 
between the red sequence and the blue cloud.

Figure~\ref{fcmd1} shows the rest-frame $U\!-\!V$ color plotted against
stellar masses for galaxies of different visual classes: Mergers,
Non-interacting Irr-1, and Non-interacting Symmetric
($\S$Section~\ref{smvc}). The visual classes of galaxies on the blue cloud
and in the red sample are shown in Table~\ref{tvccol}.  As described
in $\S$Section~\ref{sinte1}, we only consider the 20 distorted mergers listed
in Table~\ref{tmer}, and avoid the potential projection pairs without
signs of morphological distortions.  The 20 mergers are divided into
13 mergers of type 1, 3 mergers of type 2a, and 4 mergers of type 2b.
For mergers of type 2b, which are resolved into two galaxies with
separate COMBO-17 colors, we plot the average $U-V$ color of the
galaxies in the pair.

Out of the 886 visually classified systems in our sample, we find that
310/886 or $35\%\pm7\%$ lie on the blue cloud.  Out of the 20 visually
classified mergers with distortions, 16/20 or $80\%\pm 18\%$ lie on the
blue cloud (Table~\ref{tvccol}).  Conversely, out of 866
non-interacting galaxies, 294/866 or $34\%\pm 7\%$ are on the blue
cloud.  Thus, the fraction of mergers, which lie on the blue cloud
($f_{\rm merg-BC}$) is over two times larger than the fraction of
non-interacting galaxies ($f_{\rm non-int-BC}$), which lie on the blue
cloud. This implies that mergers and interacting galaxies are
preferentially blue, compared to non-interacting galaxies .

The observed higher fraction of blue galaxies among mergers compared
to non-interacting galaxies may be caused by several factors. It may
be due to enhanced levels of unobscured SF in mergers (see
$\S$Section~\ref{ssfr1}), translating to bluer colors on average. It may also
in part be the result of the mergers being part of accreted field and
group galaxies ($\S$Section~\ref{sinte3}), which are bluer than the average
cluster galaxy.  It is also relevant to ask whether we are
overestimating the fraction of blue galaxies among interacting systems
due to the visibility timescale $t_{\rm vis}$ of morphological
distortions induced by interactions being longer in bluer galaxies
than redder ones.  While this is possible, it is non-trivial to
correct for this effect because no direct unique relation exists
between $t_{\rm vis}$ and color.  As discussed in $\S~\ref{sinte1}$,
$t_{\rm vis}$ depends on the mass ratio of an interaction as well as
the gas content.  A higher gas content may result in a longer $t_{\rm
vis}$ for certain gas distributions, but it can lead to either redder
colors (e.g., enhanced level of dusty SF) or bluer colors (enhanced
level of unobscured SF), compared to interacting systems.

\subsection{SF properties of interacting galaxies}\label{ssfr1}
This work uses SFRs based on UV data from COMBO-17 (Wolf \etal 2004)
and $Spitzer$ 24$\micron$ imaging (Bell \etal 2007). The unobscured
SFR$_{\rm UV}$ is derived using the 2800\ \AA \ rest frame luminosity
($L_{\rm UV}$ = 1.5$\nu l_{\nu,2800}$) as described in Bell \etal
(2005, 2007).  The UV spectrum is dominated by continuum emission from
massive stars and provides a good estimate of the integrated SFR from
the younger stellar population in the wavelength range of 1216-3000\
\AA. \ The SFR$_{\rm IR}$ is derived using the 24$\micron$ flux to
construct the integrated IR luminosity ($L_{\rm IR}$) over
8-1000$\micron$ following the methods of Papovich \& Bell (2002).  The
total SFR is derived using identical assumptions of Kennicutt (1998)
from PEGASE assuming a 100 Myr old stellar population with constant
SFR and a Chabrier (2003) IMF:
\begin{equation}\label{sfrtot}
{\rm SFR}_{\rm UV + IR} = 9.8 \times 10^{-11}(L_{\rm IR} + 2.2L_{\rm UV}).  
\end{equation}
The factor of 2.2 on the UV luminosity accounts for light being
emitted longward of 3000\ \AA \ and shortward of 1216\ \AA \ by young
stars.  The total SFR accounts for both the dust-reprocessed (IR) and
unobscured (UV) SF.

We work with our final sample of 886 classifiable bright massive
($M_{*} \geq 10^{9} M_{\sun}$) systems, and only consider here the 20
distorted mergers in Table~\ref{tmer}.  All of these systems have
UV-based SFR from COMBO-17 observations. Of this sample, $\sim$11\%
(94/886) were not observed with $Spitzer$, $\sim$23\% (206/886) were
observed and detected at 24$\micron$, while the rest had no detection
at the $\sim$4$\sigma$ depth of 83 $\mu$Jy.

The UV-based SFR (SFR$_{\rm UV}$) versus stellar mass is plotted in
Figure~\ref{fsfruv} for all 886 systems.  The SFR$_{\rm UV}$ ranges
from $\sim$0.01 to 14 $M_{\sun}$ yr$^{-1}$.  The UV-based SFR
represents a lower limit to the total SFR for galaxies on the blue
cloud and most star-forming galaxies on the red sample. However, for
some old red galaxies, the SFR$_{\rm UV}$ may overestimate the true
SFR as the UV light from such systems may not trace massive OB stars,
but rather low to intermediate mass stars.

Figure~\ref{fsfrtot} shows the UV+IR-based SFR (SFR$_{\rm UV + IR}$),
which ranges from $\sim$0.2 to 9 $M_{\sun}$ yr$^{-1}$.  For the 206
galaxies that were observed and detected at 24$\micron$, the implied
UV+IR-based SFR is plotted as stars in the lower panel of
Figure~\ref{fsfrtot}.  For the 586 galaxies that are observed but
undetected with $Spitzer$, we use the 24$\micron$ detection limit as
an upper limit on the 24$\micron$ flux.  The corresponding upper limit
on the UV+IR-based SFR is plotted as inverted triangles in the lower
panel of Figure~\ref{fsfrtot}, and included in the calculation of the
average UV+IR-based SFR, plotted in the middle panel.

In a cluster environment, the competition between processes that
enhance the SFR and those that depress the SFR, ultimately determine
the average SFR of cluster galaxies.  The first class of processes are
strong close interactions: tidal and mergers (e.g., Toomre \& Toomre
1972), and harassment (e.g., Moore \etal 1996), which refers to the
cumulative effect of weak interactions.  The second class of processes
include ram pressure stripping of cold gas out of the galaxy by the
hot ICM (e.g., Gunn \& Gott 1972), and strangulation (e.g., Larson
\etal 1980; Balogh, Navarro \& Morris 2000).

For the few mergers (orange line) present, the average SFR$_{\rm UV}$
is typically enhanced by an average modest factor of $\sim$2 compared
to the both Non-interacting Symmetric and Irr-1 galaxies (purple,
green and black lines).  Similarly, the UV+IR-based SFR (SFR$_{\rm UV
+ IR}$; and Figure~\ref{fsfrtot}) of merging galaxies is typically
enhanced by only an average factor of $\sim$1.5 compared to the
Non-interacting Symmetric galaxies (purple line) and to all
Non-interacting galaxies (i.e Symmetric + Irr-1; black line).  We note
that a similar modest enhancement in the average SFR, by a factor of
1.5--2 is also found in mergers in the field over $z\sim$~0.24--0.80
by Jogee \etal (2008, 2009). This modest enhancement is consistent
with the theoretical predictions of di Matteo \etal (2007; see their
Figure 10), based on a recent statistical study of several hundred
simulated galaxy collisions.  Modest SFR enhancements are also seen in
galaxy pair studies in the field (Barton \etal 2000,2003; Lin \etal
2004; Ellison \etal 2008) and in mixed environments (Robaina \etal
2009; Alonso \etal 2004).

While mergers in the A901/902 clusters enhance the SFR of individual
galaxies, it is clear that they do not contribute much to the total
SFR of the cluster. We compute the SFR density of mergers to
non-interacting galaxies in the same volume of A901/902 by taking the
ratio of the total SFR in each class.  If we include only the 20
distorted mergers in Table~\ref{tmer}, we find that the contribution
of mergers to the SFR density of the clusters to be 10\%.
Alternatively, if we include all of the 36 visually classified mergers
in Table~\ref{tmer}, before accounting for false projection pairs, we
find the contribution of mergers to the SFR density to be 15\%. Thus,
we find that mergers contribute only a small fraction (between 10\%
and 15\%) of the total SFR density of the A901/902 clusters compared
to non-interacting galaxies. The small contribution of mergers to the
total cluster SFR density is likely due to the low number of mergers
in A901/902, and the fact that these mergers only cause a modest SFR
enhancement.

\section{Summary}\label{ssumm}

We present a study of the frequency, distribution, color, and SF
properties of galaxy mergers in the A901/902 supercluster at
$z\sim$~0.165, using a sample of 893 bright ($R_{\rm Vega} \le$~24)
intermediate mass ($M_{*} \geq 10^{9} M_{\sun}$) galaxies.  The sample is
complete in stellar mass down to $10^{9} M_{\sun}$.  We use $HST$ ACS
F606W data from the STAGES survey, COMBO-17, $Spitzer$ 24$\micron$,
and $XMM$-$Newton$ X-ray data. Our results are as follows:

\begin{itemize}

\item 
We visually classify the ACS F606W images of the sample galaxies into
five main visual classes: Mergers of types 1, 2a, and 2b, and
Non-Interacting systems of type Symmetric and Irr-1 ($\S$Section~\ref{smvc};
Tables~\ref{tvc1}--\ref{tcut}; Figures \ref{fvcsch}, \ref{fmer}, and
\ref{fexam3}) and obtain a final sample of 886 systems.  We classify
the systems as mergers if they show evidence of a recent or ongoing
merger of mass ratio $ >$~1/10.  Mergers of type 1 essentially appear
as an advanced merger remnant, which hosts morphological distortions
similar to those produced in simulations of mergers of mass ratio
$>$~1/10. Mergers of type 2a and 2b appear in ACS images as a very
close ($d<$ 10 kpc) overlapping pair of two galaxies. The main
difference between them is that mergers of type 2b are resolved into
two separate galaxies by the ground-based COMBO-17 data, while mergers
of type 2a are not.

We perform Monte Carlo simulations  to gauge the level of contamination 
from projection pairs. In the final analysis, we minimize the contamination 
from false projection pairs by  considering only mergers of type 1 and
morphologically distorted systems of type 2a and 2b as reliable merger
candidates.  We identify 20 morphologically distorted mergers (13 of
type 1, 3 of type 2a, and 4 of type 2b) in the sample.  The fraction
$f_{\rm merge}$ of systems with $M_{*} \geq 10^{9} M_{\sun}$, which
show evidence of a recent or ongoing merger of mass ratio $ >$~1/10 is
0.023$\pm$0.007.  The estimated fractions of likely major mergers,
likely minor mergers, and ambiguous cases are 0.01$\pm$0.004 (9/886),
0.006$\pm$0.003 (5/886), and 0.007$\pm$0.003 (6/886), respectively.
Similar merger fractions are obtained among systems with $M_{*} \ge
3.0 \times 10^{9} M_{\sun}$.

\item 
We also estimate the merger fraction using the automated CAS merger
criterion ($A>0.35$ and $A>S$; $\S$Section~\ref{scas1} and
Figure~\ref{fcas1}).  Among systems with $M_{*} \geq 10^{9} M_{\sun}$,
we find the merger fraction to be $\sim$18/886 or $0.02\pm0.006$.
While the CAS-based merger fraction may not be widely different from
the visually-based merger fraction, the two methods pick different
systems.  The CAS criterion only captures 11 of the 20 ($55\%\pm20\%$)
visually classified mergers. Furthermore, the remaining 7/18
($39\%\pm14\%$) of the systems captured by CAS turn out to be
non-interacting systems (Table~\ref{tcas}).  These contaminants tend
to be dusty or highly inclined galaxies and systems with low level
asymmetries, which are probably due to SF (Figure~\ref{fcas2}).

\item 
We compare our visually-based merger fraction in the A901/902 clusters
with those reported in other clusters and groups out to $z\sim$~0.8
($\S$Section~\ref{scomp1} and Figure~\ref{fcompa}).  The low merger fractions
among intermediate mass ($10^{9}$ to a few $\times 10^{10} M_{\sun}$)
or intermediate luminosities ($L < L^{*}$) galaxies in the A901/902
clusters and other low redshift clusters are consistent with
predictions from hierarchical models.  However, we cannot yet test the
predicted rate of evolution of the merger fraction with redshfit. Data
on the merger fraction among $L \leq L^{*}$ cluster galaxies, based on
our study and three other published studies, allow for a wide spectrum
of scenarios, ranging from no evolution to evolution by a factor of
$\sim$5 over $z\sim$~0.17 to 0.4.

\item 
Throughout this paper, we consider the core of each cluster to be at a
projected clustocentric radius of $R\le$~0.25 Mpc.  We refer to the
region between the cluster core and the virial radius as the outer
region of the clusters (0.25 Mpc~$<R \le$~1.2 Mpc).  We refer to the
region outside the virial radius (1.2 Mpc~$< R \le$~2.0 Mpc) as the
outskirt region of the clusters.  The mergers are found to lie outside
the cluster core ($\S$Section~\ref{sinte2} and Figure~\ref{fdist1}), although
the timescale for collisions and close encounters is shortest in the
core ($<$~1 Gyr). We suggest that this is due to the large velocity
dispersion (700--900 km s$^{-1}$) of galaxies in the core. Such a
dispersion makes it less likely that a close encounter between two
galaxies culminates into a merger or a disruptive interaction
associated with a large amount of tidal heating.

All of the galaxy mergers lie in the region (0.25 Mpc~$<R\le$~2 Mpc),
between the core and cluster outskirt, although the timescale for
collisions and close encounters is quite large ($\gg$~1 Gyr) in this
region. One possible explanation might be that the galaxy velocity
dispersion drops in the outer region, hence favoring mergers. However,
limited number statistics of current spectroscopic data do not allow
us to assess this possibility. Another possible scenario is that the
interacting galaxies in the outer region and cluster outskirt are part
of groups or field galaxies that are being accreted along cosmological
filaments by the A901/902 clusters. We find that our estimated number
density and fraction of mergers to be similar to that predicted at
typical group overdensities in $N$-body simulations of groups and
field galaxies accreting into the A901/902 clusters
(Figure~\ref{felco1}).  This suggests the ongoing growth of the
clusters via accretion of group and field galaxies.  The
preferentially blue color of the mergers (see below) also supports
this scenario.

\item 
Out of the 20 morphologically distorted merger remnants and merging
pairs in the sample, 16 lie on the blue cloud and 4 are in the red
sample ($\S$Section~\ref{scolor1} and Figure~\ref{fcmd1}).  The fraction of
mergers, which lie on the blue cloud is 80\%$\pm 18\%$ (16/20).  This is
over a factor of 2 higher than the fraction (294/866 or $34\%\pm 7\%$)
of non-interacting galaxies, which lie on the blue cloud, suggesting
that mergers are preferentially blue compared to non-interacting
galaxies.  This effect may be due to the enhanced recent SF in mergers
and/or due to the possibility that the mergers are part of group and
field galaxies, which are accreting into the cluster.

\item 
Among intermediate mass $M_{*} \geq 10^{9} M_{\sun}$ systems, the
average SFR$_{\rm UV}$ and SFR$_{\rm UV + IR}$ of mergers are
typically enhanced by only a modest factor of $\sim$2
(Figure~\ref{fsfruv}) and $\sim$1.5 (Figure~\ref{fsfrtot}),
respectively, compared to the non-interacting galaxies (i.e.,
Symmetric + Irr-1).  This modest enhancement is consistent with the
theoretical predictions of di Matteo \etal (2007), based on a recent
statistical study of several hundred simulated galaxy collisions.  Our
results of a modest enhancement and a low merger fraction culminate in
our finding that mergers contribute only  a small fraction 
(between 10\% and 15\%)  of the total SFR density of the A901/902 clusters.
\end{itemize}


A.H. and S.J. gratefully acknowledge support from NSF grant
AST-0607748, LTSA grant NAG5-13063, as well as programs HST-GO-10395,
HST-GO-10861, and HST GO-11082, which were supported by NASA through a
grant from the Space Telescope Science Institute, which is operated by
the Association of Universities for Research in Astronomy,
Incorporated, under NASA contract NAS5-26555. We thank Christopher
Conselice for his assistance with CAS.  E.vK. and M.B. were supported
by the Austrian Science Foundation FWF under grant P18416,C.Y.P. by
STScI and NRC-HIA Fellowship programmes, C.H. by European Commission
Programme Sixth Framework Marie Curie Outgoing International
Fellowship under contract MOIF-CT-2006-21891 and a CITA National
fellowship, M.E.G. and C.W. by an STFC Advanced Fellowship, E.F.B. and
K.J. by the DFG's Emmy Noether Programme, A.B. by the DLR (50 OR
0404), S.F.S. by the Spanish MEC grants AYA2005-09413-C02-02 and the
PAI of the Junta de Andaluc as research group FQM322, L.V.W. by NSERC,
CIfAR and CFI, and D.H.M. by NASA under LTSA Grant NAG5-13102. The
STAGES team thanks Hans-Walter Rix for providing essential support
contributing to the success of the STAGES project. This research made
use of NASA's Astrophysics Data System and NASA/IPAC Extragalactic
Database.



}
{}
                                                                                

\begin{deluxetable}{lcc}
\tabletypesize{\scriptsize}
\tablewidth{0pt}
\tablecaption{Visual Classification Results for the sample of  bright
 intermediate-mass ($M_{\rm *} \geq 10^{9} M_{\sun}$) galaxies}
\tablehead{
\colhead{Visual Class} &
\colhead{$N_{\rm VC}$} &
\colhead{$f_{\rm VC}$} \\
\colhead{(1)} &
\colhead{(2)} &
\colhead{(3)} \\
}
\startdata
Mergers of Type 1 + 2a+ 2b & 13+3+4 = 20 & 0.023$\pm$0.007\\
Non-interacting Irr-1 & 123 & 0.14$\pm$0.03\\ 
Non-interacting Symmetric & 743 & 0.80$\pm$0.16 \\ 
Unclassifiable &6& -\\ \enddata
\tablecomments{ The results of visual classification (see
$\S$Section~\ref{smvc} for details) for the sample of  
bright  intermediate-mass ($M_{\rm *} \geq 10^{9} M_{\sun}$) systems. Columns: (1) Visual class (VC); (2) $N_{\rm VC}$ = Number of systems with
a given VC.; (3) $f_{\rm VC}$ = Fraction of systems with a given VC.
}
\label{tvc1}
\end{deluxetable}

\clearpage 

\begin{deluxetable}{ccccccc}
\tabletypesize{\scriptsize}
\tablewidth{0pt}
\tablecaption{Visual Classification of Mergers in the Sample of Bright
  Intermediate-mass ($M_{\rm *} \geq 10^{9} M_{\sun}$) galaxies}
\tablehead{
\colhead{Merger ID} &
\colhead{Merger Mass} &
\colhead{$M_{1}$}&
\colhead{$M_{2}$}&
\colhead{$M_{1}$/$M_{2}$}&
\colhead{$L_{1}$/$L_{2}$}&
\colhead{Major, Minor,}\\
\colhead{ }&
\colhead{}&
\colhead{for Type 2b} &
\colhead{for Type 2b} &
\colhead{for Type 2b} &
\colhead{for Type 2a} &
\colhead{or Ambiguous}\\
\colhead{} &
\colhead{($10^{9} M_{\sun}$)}&
\colhead{($10^{9} M_{\sun}$)}&
\colhead{($10^{9} M_{\sun}$)}&
\colhead{ } &
\colhead{} &
\colhead{}\\
\colhead{(1)} &
\colhead{(2)} &
\colhead{(3)} &
\colhead{(4)} &
\colhead{(5)} &
\colhead{(6)} &
\colhead{(7)} \\
}
\startdata 
&&&&&&\\
\multicolumn{7}{c}{\bf Mergers of Type 1} \\
1$^{\star}$ &		1.22  &&&&&   Ambiguous\\
2$^{\star}$ &		1.31  &&&&&  Ambiguous\\
3$^{\star}$ &		1.62  &&&&&  Ambiguous\\
4$^{\star}$ &		1.64   &&&&&  Ambiguous\\
5$^{\star}$ &		1.64   &&&&&  Ambiguous\\
6$^{\star}$ &		2.73  &&&&&  Ambiguous\\
7$^{\star}$ &		6.66  &&&&&  Major\\
8$^{\star}$ &		8.06  &&&&&  Minor\\
9$^{\star}$   &     	23.3  &&&&&  Minor\\
10$^{\star}$ &		24.4 &&&&&  Major\\
11$^{\star}$ &		46.0  &&&&&  Major\\
12$^{\star}$ &		122.1 &&&&&  Major\\
13$^{\star}$ &		191.5  &&&&&  Minor\\\hline
&&&&&&\\
\multicolumn{7}{c}{\bf Potential Mergers of Type 2a} \\
14 &	2.32 &&&& 6.9 & Minor\\
15 &	5.25 &&&& 6.0 & Minor\\
16$^{\star}$ &	13.0  &&&& 1.4&Major\\
17$^{\star}$ &	36.5 &&&& 5.0& Minor	\\			    
18$^{\star}$ &	281.5 &&&& 1.1& Major\\\hline
&&&&&&\\
\multicolumn{7}{c}{\bf Potential Mergers of Type 2b} \\
19 & 1.52 &1.32 & 0.20 & 6.60 & & Minor\\
20$^{\star}$ & 1.95  &1.12 & 0.83 & 1.35 & & Major\\
21 & 3.24 &2.48 & 0.76 & 3.26 & & Major\\
22&	7.90& 5.34&    2.56&2.09&& Major\\
23 & 14.6 &13.3 & 1.33 & 10. & & Minor\\
24$^{\star}$& 20.0 &15.1&  4.94&3.06&& Major\\
25&21.0&14.3&   6.73&2.12&& Major\\
26&33.6&24.1	&    9.45& 2.55&& Major\\
27& 43.5 &35.6 & 7.86 & 4.53 & & Minor\\
28& 44.3 & 26.1 & 18.2 & 1.43 & & Major\\
29& 58.9 & 37.4 & 21.5 & 1.74 & & Major\\
30&60.5&52.2&  8.34& 6.26&& Minor\\
31&79.2&71.8&  7.38& 9.73&& Minor\\	    
32&110.6&83.1& 27.5&3.02&& Major\\
33 & 126.4 & 113.8 & 12.6 & 9.03 & & Minor\\
34$^{\star}$&	154.6  &	117.0& 37.6&   	3.11&& Major\\
35$^{\star}$&	202.6  &	169.0& 16.6&   	10.18&& Minor\\
36& 238.6 &214.6 & 24.0 & 8.94 & & Minor\\
\enddata
\tablecomments{The table shows the visually classified
mergers, which are identified in the sample of 893 
bright ($R_{\rm Vega} \le$~24)  
intermediate-mass ($M_{\rm *} \geq 10^{9} M_{\sun}$) galaxies.  The
mergers are sub-divided into three groups: mergers of type 1 and {\it potential}
mergers of type  2a and 2b, according to the criteria 
outlined in $\S$Section~\ref{smvc}.  Table Columns: (1)~Numerical identifier corresponding to 
the potential mergers in Figure~\ref{fmer}. Note that the 
very close ($d<$ 10 kpc) pairs listed as potential mergers of Type
2a and 2b in this table includes both real and projection pairs. The likely 
real pairs are marked with a star in Column 1 and contain  
at least one galaxy with morphological distortions indicative 
of a galaxy-galaxy interaction. In the final analysis, only 
the 13 distorted mergers of Type 1 and the 7 distorted mergers 
of types 2a and 2b are used; 
(2)~The mass of the merger.  For pairs of type 2b,
which are resolved into two galaxies by COMBO-17 data, the mass cited
is the sum ($M_{1} + M_{2}$) of the pair members;
(3)~For mergers of Type 2b, the mass $M_{1}$ of one galaxy in the
pair; 
(4)~For mergers of Type 2b, the mass $M_{2}$ of the second
galaxy in the pair; 
(5)~For mergers of Type 2b, the mass ratio
($M_{1}$/$M_{2}$); 
(6)~For mergers of Type 2a, the luminosity ratio
($L_{1}$/$L_{2}$), measured from ACS images; 
(7)~The classification of the merger into major merger, minor merger,
or ambiguous `major or minor', according to the criteria outlined in
$\S$Section~\ref{smvc}. }
\label{tmer}
\end{deluxetable}

\clearpage
\begin{deluxetable}{cccccccccc}
\tabletypesize{\scriptsize}
\tablewidth{0pt}
\tablecaption{Merger fraction for different mass cuts}
\tablehead{
\colhead{$M_{\rm cut}$ } &
\colhead{$N_{\rm tot}$ } &
\colhead{$N_{\rm 1}$ }&
\colhead{$N_{\rm 2a}$ (all) }&
\colhead{$N_{\rm 2b}$ (all) } &
\colhead{$N_{\rm merge}$ (all)}  &
\colhead{$f_{\rm merge}$ (all)} &
\colhead{$f_{\rm major}$ (all)} &
\colhead{$f_{\rm minor}$ (all)}&
\colhead{$f_{\rm ambig}$ (all)}\\
\colhead{($M_{\sun}$)} &
\colhead{} &
\colhead{} & 
\colhead{} &
\colhead{} & 
\colhead{} &
\colhead{} &
\colhead{} &
\colhead{} &
\colhead{} \\
\colhead{(1)} &
\colhead{(2)} &
\colhead{(3)} & 
\colhead{(4)} &
\colhead{(5)} & 
\colhead{(6)} &
\colhead{(7)} &
\colhead{(8)} &
\colhead{(9)} &
\colhead{(10)}\\ 
}
\startdata 
 $1\times 10^{9}$  &  886 &  13 &   3(5)&  4(18) & 20(36)&   0.023$\pm$0.007 (0.041$\pm$0.01)&
 9/886 (16/886) & 5/886 (14/886) & 6/886 (6/886)\\ 
 $3 \times 10^{9}$ & 609 &  7  &    3(4)&   3(16)& 13(27)& 0.021$\pm$0.007  (0.033$\pm$0.01)&
 8/609 (15/609) & 3/609 (12/609) & 0 (0)\\ 
\enddata
\tablecomments{
The merger fraction $f_{\rm merge}$  among systems above a mass cut ($M_{\rm cut}$) is computed as ($N_{\rm merge}/N_{\rm tot}$), where $N_{\rm merge}$ is the number of 
galaxies with $M_{*} \ge M_{\rm cut}$ involved in major and minor mergers, 
and $N_{\rm tot}$ is the total number of systems with $M_{*} \ge M_{\rm cut}$. 
This table shows the value of  $f_{\rm merge}$ for different mass cuts.
Columns:
(1) Mass cut  $M_{\rm cut}$ in units of $M_{\sun}$;
(2) The total number $N_{\rm tot}$ of systems with $M_{*} \ge M_{\rm cut}$;
(3) Number $N_{\rm 1}$ of mergers of type 1 for this mass cut;
(4) Number $N_{\rm 2a}$ of mergers of type 2a for this mass cut.; 
(5)  Number $N_{\rm 2b}$ of  mergers of type 2b for this mass cut;
(6)  $N_{\rm merge}$, the total number of mergers of types 1, 2a, and 2b  for this mass cut; 
(7)  The merger fraction $f_{\rm merge}$, computed as ($N_{\rm merge}/N_{\rm tot}$);
(8)  The major  merger fraction, based on the classification in column 8 of Table \ref{tmer};
(9)  Same as (8), but for the minor  merger fraction; 
(10) Same as (8), but for the fraction of ambiguous mergers, which
could be either major or minor mergers. In Columns 4--10,  the first number without brackets apply to the case  we only consider the 13 type 1 mergers 
and the subset of  7 reliable type 2a and 2b mergers, which are likely to be real pairs as 
they contain at least one member galaxy with morphological  distortions  
indicative of a galaxy-galaxy interaction.  The value in bracket apply 
to the case where we include  the 13 type 1 mergers and all the 23
mergers of type 2a and 2b listed in Table~\ref{tmer}, even those which are likely to be projection pairs and contain 
galaxies without morphological distortions.
}
\label{tcut}
\end{deluxetable}

\begin{deluxetable}{lcc}
\tabletypesize{\scriptsize}
\tablewidth{0pt}
\tablecaption{Visual Classes of Galaxies Satisfying the  CAS Criterion ($A>0.35$ and $A>S$)}
\tablehead{
\colhead{Visual Class} &
\colhead{$N_{\rm CAS-VC}$} &
\colhead{$f_{\rm CAS-VC}$} \\
\colhead{(1)} &
\colhead{(2)} &
\colhead{(3)} \\
}
\startdata
All Visual Types & 18& 1.00\\\hline
Merger &11 & 11/18=0.61$\pm$0.2\\
Non-interacting Irr-1& 5& 5/18=0.28$\pm$0.1\\
Non-interacting Symmetric &2 & 2/18=0.11$\pm$0.08\\
\enddata 
\tablecomments{ The table shows the visual classes (VC) of  the 
galaxies, which satisfy the CAS criterion ($A>0.35$ and $A>S$), among
the sample of intermediate-mass  ($M_{\rm *} \geq 10^{9} M_{\sun}$) systems. Columns: 
(1) Visual Class (VC) of galaxies satisfying the CAS criterion;
(2) $N_{\rm CAS-VC}$: the number of galaxies with this visual class satisfying 
the CAS criterion;
(3) $f_{\rm CAS-VC}$: the fraction of all the galaxies satisfying the CAS criterion, which have this visual class.
}
\label{tcas}
\end{deluxetable}

\begin{deluxetable}{lccl}
\tabletypesize{\scriptsize} 
\tablewidth{0pt} \tablecaption{Comparison of Projected Galaxy Number
  Densities $n$ and $\sigma_{gal}$}
\tablehead{ 
\colhead {System} & 
\colhead{Core} & 
\colhead{Outskirt}&
\colhead {$\sigma_{\rm gal}$}\\ 
\colhead {} & 
\colhead{$R\le$~0.25
Mpc} & 
\colhead{$R >$~1.2 Mpc} & 
\colhead {} \\ 
\colhead {} & 
\colhead {(galaxies Mpc$^{-3}$)} & 
\colhead {(galaxies Mpc$^{-3}$)} & 
\colhead {(km s$^{-1}$)} \\
}  
\startdata 
Virgo$^{a}$ & 360 & 75 & 400-750\\
Coma$^{b}$ & 10000 & 400 & $\sim900$\\ 
A901a$^{c}$ & 1600 & 60 & 890$\pm182$\\ 
A901b$^{c}$ & 1100 & 55 & 1189$\pm266$\\ 
A902$^{c}$ & 1000 & 40 & 792$\pm176$\\ 
Groups$^{d}$ & $\sim$0.01 & & $\sim$100\\
\enddata 
\tablecomments{ This table compares the projected galaxy
number densities $n$ and galaxy velocity dispersions $\sigma_{gal}$ in
the A901a, A901b, and A902 clusters to those of groups and other
clusters, such as Coma and Virgo. For the A901 clusters, the value of
$n$ is computed using the sample of 
bright ($R_{\rm Vega} \le$~24)  
intermediate-mass ($M_{*} \geq 10^{9} M_{\sun}$) galaxies.  References
for the values cited are: a~=~Binggeli,Tammann, \& Sandage (1987);
b~=~The \& White (1986); c = number densities are from this work, and 
the average local velocity dispersions are from Gray \etal (in preparation); 
d~=~ Tago \etal (2008)}
\label{tcompa}
\end{deluxetable}

\begin{deluxetable}{lccc}
\tabletypesize{\scriptsize} 
\tablewidth{0pt} 
\tablecaption{Properties in Core, Outer Region, and Outskirt of A901/902 Clusters}
\tablehead{
\colhead{} &
\colhead{Cluster Core} &
\colhead{Cluster Outer Region} &
\colhead{Outskirt} \\
\colhead{} &
\colhead {($R\le$~0.25 Mpc)} &
\colhead {(0.25~Mpc~$< R \le$~1.2 Mpc)} &
\colhead {(1.2 Mpc~$< R \le$~2.0 Mpc)} \\
}
\startdata 
(1)  $N$  &85  &  533&  195\\ 
(2) Volume $V$ (Mpc${^3}$) & 0.065 & 7.17& 26.3\\
(3) $n$ (gal Mpc$^{-3}$)& 1307 &  74.3 &7.41\\
(4) $\langle B\!-\!V \rangle$ & 0.839 &  0.764 &0.739\\
(5) $\langle U\!-\!V \rangle$ & 1.25 &  1.08 & 1.03\\
(6) $t_{\rm coll}$ (Gyr)& 0.765 & 13.5 & 135\\
\hline
(7)  $N_{\rm non-int}$  &84  &  519&  192\\ 
(8) $N_{\rm merger}$   &  0 & 15& 3 \\
(9) $n_{\rm merger}$ (merger Mpc$^{-3}$) & 0   &  2.09 & 0.11\\
(10) $f_{\rm merger}$  & 0&  0.028&  0.015\\
(11) $N_{\rm major-merger}$ &  0 &  8 & 0\\
(12) $n_{\rm major-merger}$ (merger Mpc$^{-3}$)&   0 & 1.12 &0\\
(13) $f_{\rm major-merger}$& 0 & 0.015&  0.005\\
(14) $N_{\rm minor-merger}$ &  0 &  5 & 0\\
(15) $n_{\rm minor-merger}$ (merger Mpc$^{-3}$)&   0 & 0.70 &0\\
(16) $f_{\rm minor-merger}$& 0 & 0.009&  0\\\hline 

(17) $N_{\rm non-int-BC}$ & 7 & 178& 81\\ 
(18) $f_{\rm non-int-BC}$ & 0.08 &0.34 & 0.42\\ 
(19) $N_{\rm merger-BC}$ & 0 & 11 & 3\\ 
(20) $f_{\rm merger-BC}$ & 0 & 0.73 & 1.0\\\hline 
\enddata 
\tablecomments{ This
  table shows galaxy properties and timescales in three different
  regions of the cluster: the core ($R\le$~0.25 Mpc), the outer region
  (0.25~Mpc~$< R \le$~1.2 Mpc) between the core and the cluster, and
  the outskirt region (1.2 Mpc~$< R \le$~2.0 Mpc). 
The values are computed based on the sample of 
bright ($R_{\rm Vega} \le$~24)  
intermediate-mass  ($M_{*} \geq 10^{9}  M_{\sun}$) systems.
The rows are : 
(1)~Total number of galaxies; 
(2)~Volume of region; 
(3)~Projected number density of galaxies; 
(4)~Mean $B-V$ rest frame color; 
(5)~Mean $U-V$ rest frame color; 
(6)~Timescale for collisions or close encounters; 
(7)~Number of non-interacting galaxies;
(8)~Number of mergers from Table~\ref{tmer};
(9)~Projected number density of mergers;
(10)~Fraction of mergers;
(11)-(13): same as (8)-(10), but for major mergers; 
(14)-(16): same as (8)-(10), but for minor mergers; 
(17)~Number of non-interacting galaxies on blue cloud; 
(18)~Fraction of non-interacting galaxies on blue cloud;
(19)~Number of  mergers on blue cloud; 
(20)~Fraction of mergers on blue cloud.
}
\label{t3rad}
\end{deluxetable}

\begin{deluxetable}{lcccc}
\tabletypesize{\scriptsize} 
\tablewidth{0pt} 
\tablecaption{Visual Classes of Galaxies on the Blue Cloud and Red Sample}
\tablehead{
\colhead{Visual Class}&
\colhead{$N_{\rm blue}$}&
\colhead{$f_{\rm blue}$}&
\colhead{$N_{\rm red}$}&
\colhead{$f_{\rm red}$} \\
\colhead{(1)}&
\colhead{(2)}&
\colhead{(3)}&
\colhead{(4)}&
\colhead{(5)}\\
}
\startdata
(1)  All                    &310  &  1.00                    & 576 & 1.00 \\
(2) Mergers                 & 16  &  16/310~=~0.052$\pm$0.02    & 4   & 4/576~=~0.007$\pm$ 0.004 \\
(2a) Likely Major Mergers   & 6   &  6/310~=~0.02 $\pm$ 0.009   &3    & 3/576~=~0.005$\pm$ 0.003   \\
(2b) Likely Minor Mergers   & 4   &  4/310~=~0.01 $\pm$ 0.007   & 1   & 1/576~=~0.002$\pm$ 0.002   \\
(2c) Ambiguous Major/Minor  & 6   &  6/310~=~0.02  $\pm$ 0.009  &  0  &  0 \\
(3) Non-interacting Irr-1      &105    &105/310~=~0.34$\pm$ 0.07 & 18 & 18/576~=~0.033$\pm$ 0.01 \\
(4) Non-interacting Symmetric  & 189 & 189/310~=~61$\pm$13   & 554 & 554/576~=~0.96$\pm$19 \\
\enddata
\tablecomments{ 
Results are shown for the sample of 886 
bright ($R_{\rm Vega} \le$~24)  
intermediate-mass ($M_{\rm *} \geq 10^{9} M_{\sun}$)  
visually classifiable systems. 
The table shows how galaxies in the blue cloud and red sample 
are split among the different visual classes (Mergers,  Non-interacting Irr-1, 
and Non-interacting Symmetric)  discussed in $\S$Section~\ref{scolor1}.  Columns: 
(1) Visual Classes: These include  Mergers,  Non-interacting Irr-1, 
Non-interacting Symmetric, listed in rows 2, 3, and 4.
We only consider  the 20 morphologically distorted mergers listed in 
Table~\ref{tmer}. These include  13 mergers of type 1, 3 mergers of 
type 2a, and  4 mergers of type 2b.  
Rows 2a, 2b, and 2c show how the mergers are  split into likely 
major mergers, likely minor mergers, and ambiguous cases that could 
be either ($\S$Section~\ref{smvc}). 
(2) $N_{\rm blue}$: Number of galaxies on the blue cloud;
(3) $f_{\rm blue}$: Fraction of blue cloud galaxies, which belong to  a given visual class;
(4) $N_{\rm red}$: Number of galaxies in the red sample;
(5) $f_{\rm red}$: Fraction of red sample galaxies,which belong to  a given visual class.
}
 \label{tvccol}
\end{deluxetable}


\clearpage
\begin{figure}
\epsscale{.9}
\plotone{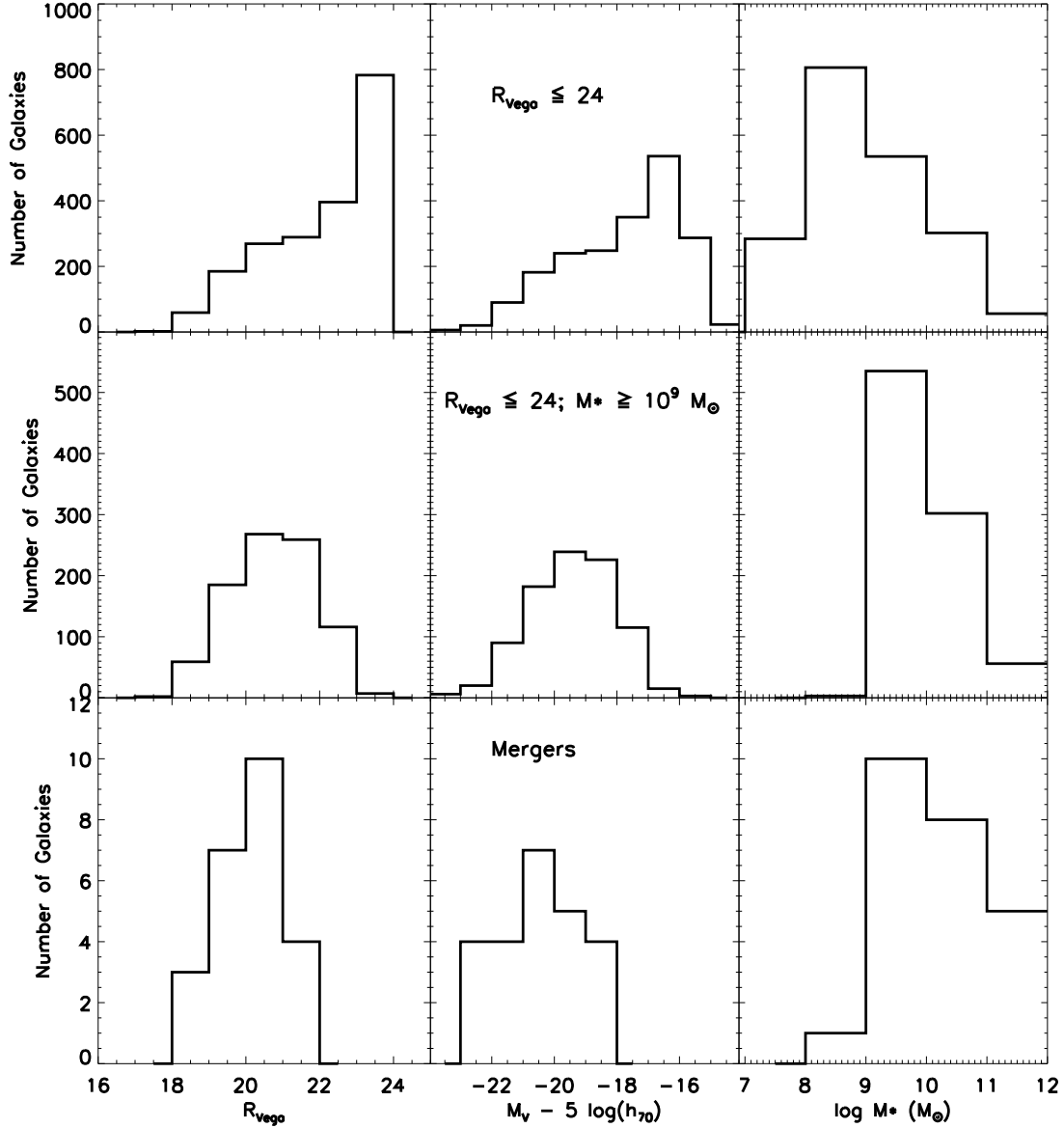}
\caption{{\bf Top panel:} Apparent $R$--band magnitude ($R_{\rm
Vega}$), absolute visual magnitude ($M_{\rm V}$), and stellar mass
($M_{*}$) distributions of the full A901/902 supercluster sample
($R_{\rm Vega} \leq 24$; 1983 galaxies). {\bf Middle panel:} Same as
the top panel, but showing the properties of the sample of 893
intermediate mass galaxies ($M_{*} \geq 10^{9} M_{\sun}$).  {\bf
Bottom panel:} Same as the middle panel, but only showing the
properties of the 24 galaxies, which are part of the 20
morphologically distorted mergers identified among the sample (see
$\S$Section~\ref{smvc} and Table~\ref{tmer} for details). These 20 mergers
are labeled with a starred identification number in Table~\ref{tmer}
and are Cases 1--13, 16--18, 20, 24, 34, 35 in Table~\ref{tmer}.  For
the four mergers of type 2b (which are close pairs resolved by COMBO-17)
we plot the properties of the eight individual galaxies making up the
pair, thus resulting in a total of 24 galaxies. Each pair has a galaxy
of mass $M_{*} \geq 10^{9} M_{\sun}$ and a second galaxy that may be
of any mass.  }
\label{flumf}
\end{figure}

\clearpage
\begin{figure}
\vspace{-1.8cm}
\epsscale{.90}
\plotone{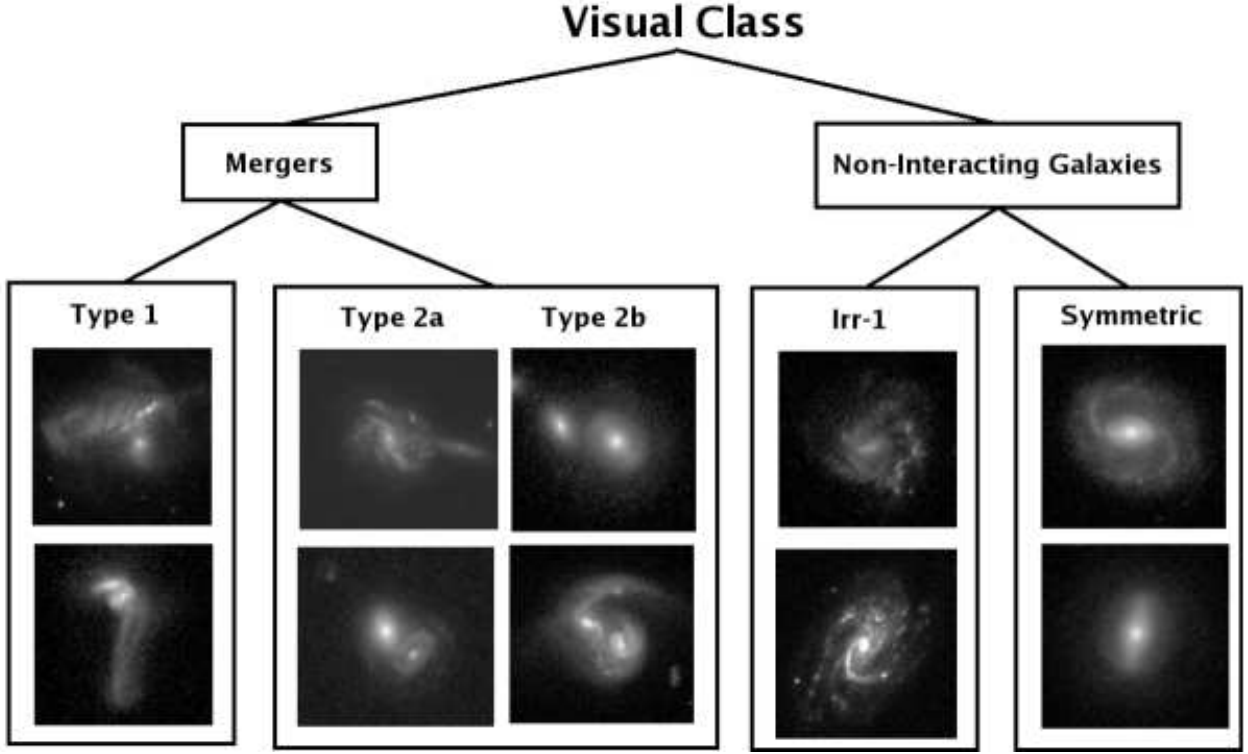}
\caption{In our visual classification scheme ($\S$Section~\ref{smvc}),
systems are classified as Mergers, or Non-interacting.  The mergers
are sub-divided into three groups called Type 1, Type 2a, and Type 2b.
Mergers of Type 1 appear, in the ACS F606W image, as a single
morphologically distorted remnant, rather than two individually
recognizable galaxies.  The remnant hosts strong {\it externally
triggered} morphological distortions similar to those seen in
simulations of mergers of mass ratio $>$~1/10, such as tidal tails,
shells, ripples, warps, strongly asymmetric tidal debris, double
nuclei inside a common envelope, or a `train-wreck' morphology. In
contrast, mergers of type 2a and 2b appear in ACS images as a very
close ($d<$ 10 kpc) overlapping pair of two recognizable galaxies.
Mergers of type 2b are resolved into two separate galaxies by the
ground-based COMBO-17 data, while mergers of type 2a are not.  Since
some of these type 2a and 2b systems could be due to chance
line-of-sight superposition, we conservatively consider for the final
analysis only those pairs where at least one member is morphologically
distorted (see Table~\ref{tmer} and Figure~\ref{fmer} for details).
Non-interacting systems are sub-divided into Irr-1 and Symmetric
systems.  Irr-1 exhibit { \it internally triggered} asymmetries, due
to SF typically on scales of a few hundred parsecs.  Symmetric systems
include galaxies, which are relatively undistorted and are not part of
the very close pairs that constitute the mergers of Types 2a and 2b.}
\label{fvcsch}
\end{figure}

\clearpage
\begin{figure}
\vspace{-1.8cm}
\epsscale{.6}
\plotone{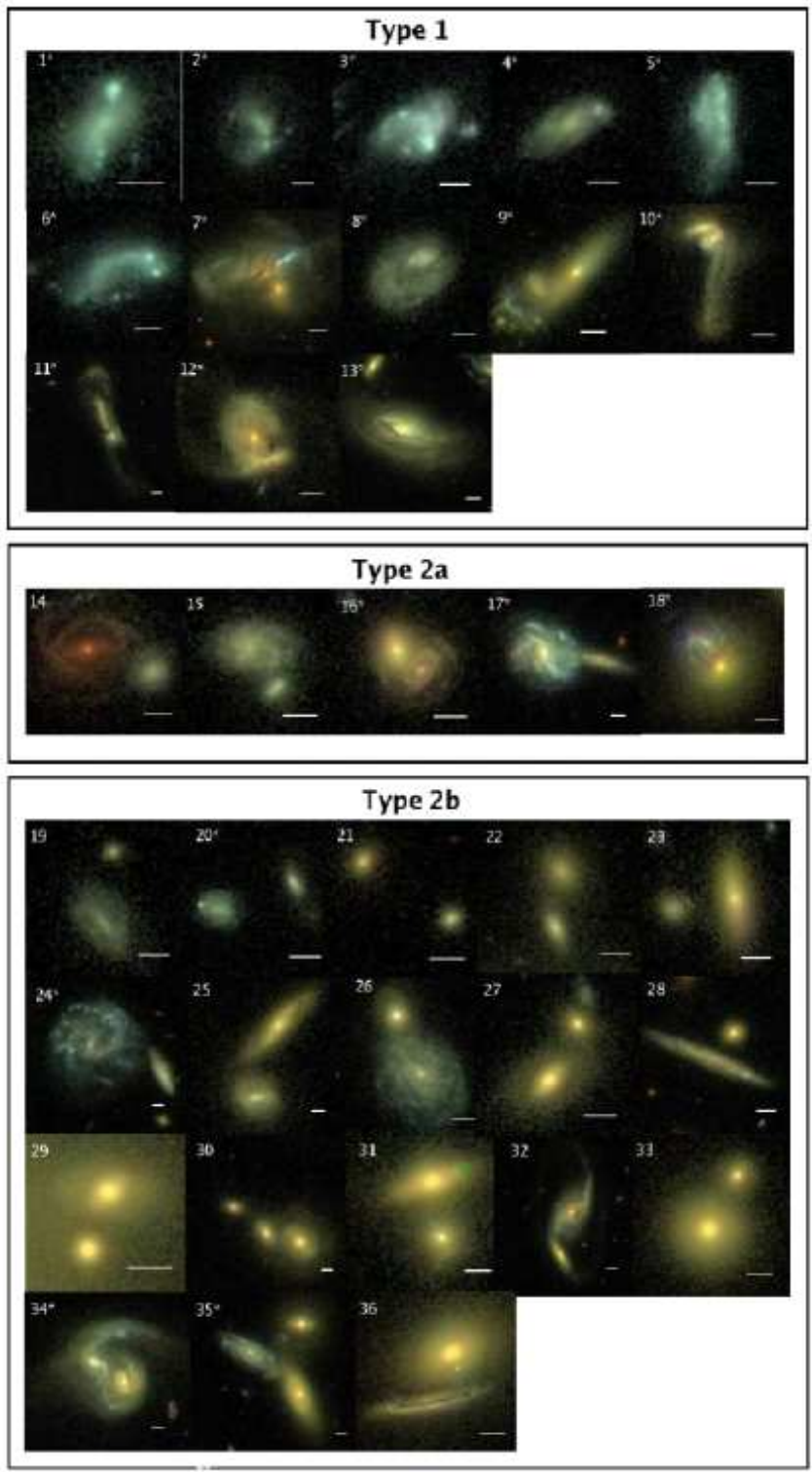}
\caption{ The top, middle, and lower panels show the ACS images of
systems visually classified as mergers of type 1 and {\it potential}
mergers of type 2a and 2b, respectively (see $\S$Section~\ref{smvc} for details).  White
horizontal bars denote a scale of $1\arcsec$ or 2.8 kpc at
$z\sim$~0.165.  The properties of the mergers are listed in
Table~\ref{tmer}.  {\bf Top panel}: The 13 distorted remnants
classified as type 1 mergers hosts strong {\it externally triggered}
morphological distortions similar to those seen in simulations of
mergers of mass ratio $>$~1/10, such as tidal tails (e.g., Cases 10,
11), shells, ripples, warps, asymmetric tidal debris and distortions
(e.g., Cases 5,6,8,9), double nuclei inside a common envelope, or a
`train-wreck' morphology (Cases 7, 10, 11, 12).  The type 1 mergers
can be divided into major mergers (Cases 7, 10, 11, 12), minor mergers
(Cases 8,9,13), and ambiguous `major or minor' mergers (Cases 1-6) as
described in $\S$Section~\ref{smvc}.  {\bf Middle and Lower panels}: The 23
potential mergers of type 2a and 2b shown here include both real and
projection very close ($d<$ 10 kpc) pairs.  The 7 likely real pairs
(Cases 16--18, 20, 24, 34, 35) are denoted with a starred
identification number and contain at least one galaxy with
morphological distortions indicative of a galaxy-galaxy interaction.
In the final analysis, only the 13 distorted mergers of Type 1 and the
7 distorted mergers of types 2a and 2b are used Type 2b pairs are
resolved by COMBO-17 data into two galaxies with separate redshifts
and stellar masses, and their stellar mass ratio (listed in
Table~\ref{tmer}) is used to divide Type 2b mergers into major mergers
(Cases 20--22, 24--26, 28, 29, 23, 34) and minor mergers (Cases 19,
23, 27, 30, 31, 33, 35, 36).  For the Type 2a pairs, which are
unresolved by COMBO-17, the ACS-based luminosity ratio (listed in
Table~\ref{tmer}) of the pair members is used to divide them into
major (Cases 16, 18) and minor (Cases 14, 15, 17) mergers. }
\label{fmer}
\end{figure}

\clearpage
\begin{figure}
\epsscale{0.9}
\plotone{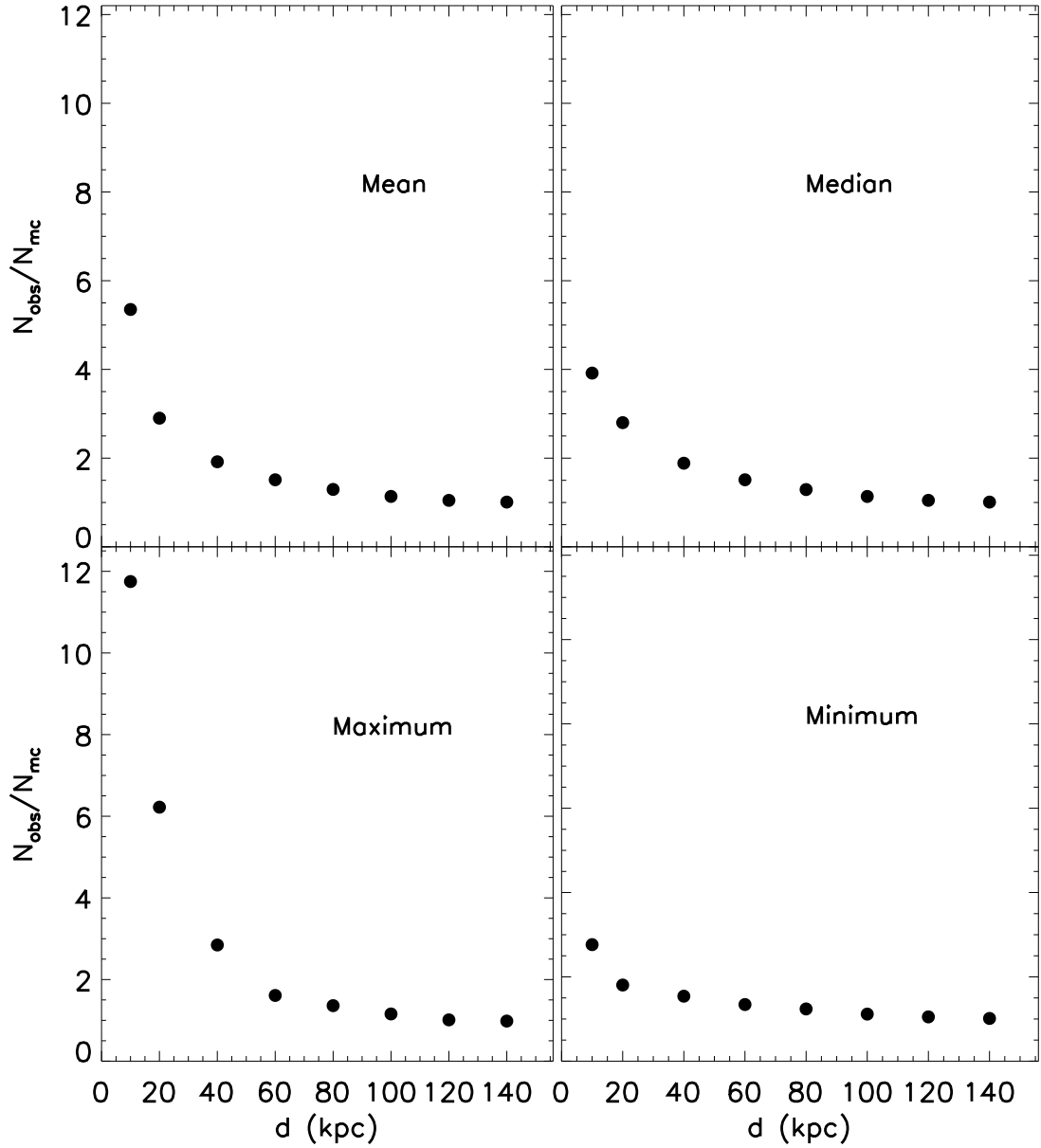}
\caption{ The ratio ($N_{\rm obs}/N_{\rm mc}$) for pairs of different
projected separations $d$, where $N_{\rm obs}$ and $N_{\rm mc}$ are
the number of pairs measured in the observations and random pairs from
the Monte Carlo simulations, respectively.  See $\S$Section~\ref{smvc}
for details.  The four panels represent the mean, median, minimum and
maximum distribution as labeled. At smaller separations of $d < 50$
kpc, ($N_{\rm obs}/N_{\rm mc}$) is above 1, suggesting that random
chance superposition cannot fully account for all the observed pairs.
}
\label{fmc}
\end{figure}

\clearpage
\begin{figure}
\epsscale{.80}
\plotone{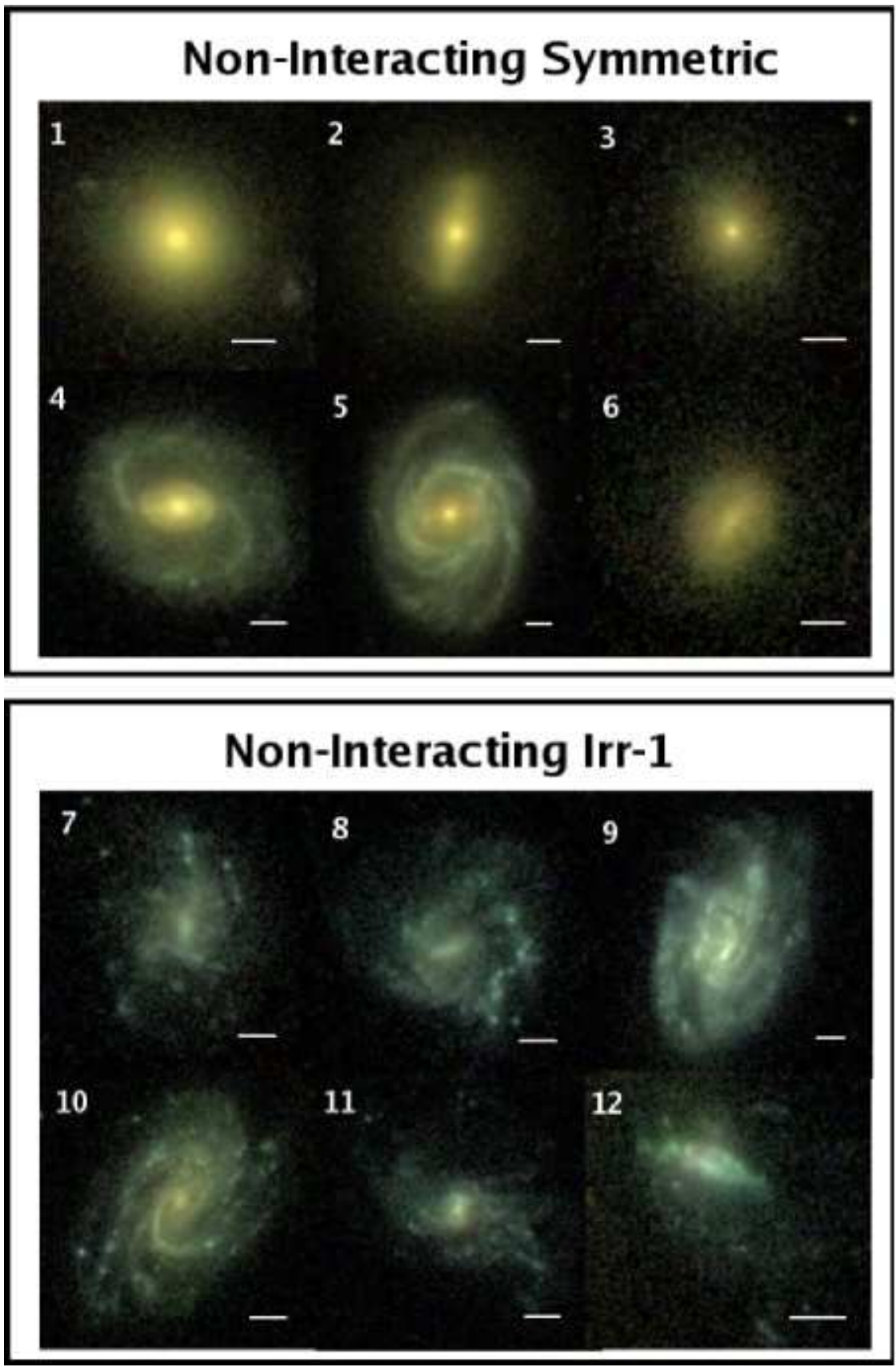}
\caption{ACS F606W images of some of the systems visually classified
as Non-interacting Symmetric (Cases 1-6) and Non-interacting Irr-1
(Cases 7-12). See $\S$Section~\ref{smvc} for details.}
\label{fexam3}
\end{figure}

\clearpage
\begin{figure}
\epsscale{.850}
\plotone{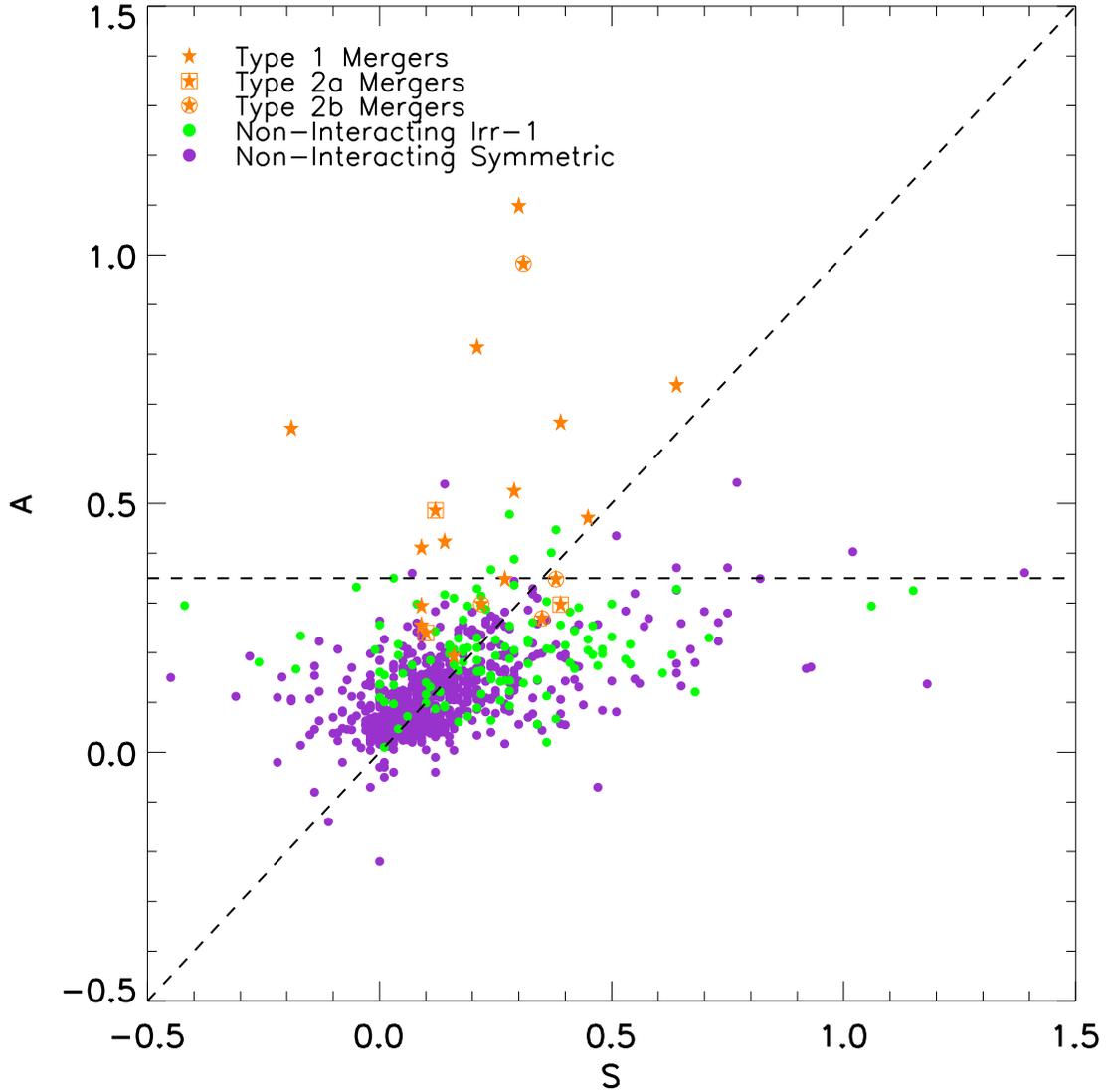}
\caption{ The results of running CAS on the sample of bright ($R_{\rm
Vega} \le$~24) intermediate-mass ($M_{*} \geq10^{9}$ $M_{\sun}$)
systems are shown.  Galaxies with different visual classes (Merger,
Non-interacting Irr-1, and Non-interacting Symmetric) are shown in the
CAS $A$ vs. $S$ plane. We only plot here the 20 distorted mergers
of types 1, 2a, and 2b in Table~\ref{tmer}.  The 18 galaxies
satisfying the CAS merger criterion ($A > S$ and $A>$~0.35) lie on the
upper left corner of the diagram.  The CAS merger criterion recovers
11 of the 20 ($61\%\pm20\%$) of the galaxies visually classified as a
merger.  Furthermore, there is a significant level of contamination:
7/18 ($39\%\pm14\%$) of the systems picked up by the CAS criterion are
visually classified as Non-interacting Irr-1 and Non-interacting
Symmetric.}
\label{fcas1}
\end{figure}

\clearpage
\begin{figure}
\epsscale{.850}
\plotone{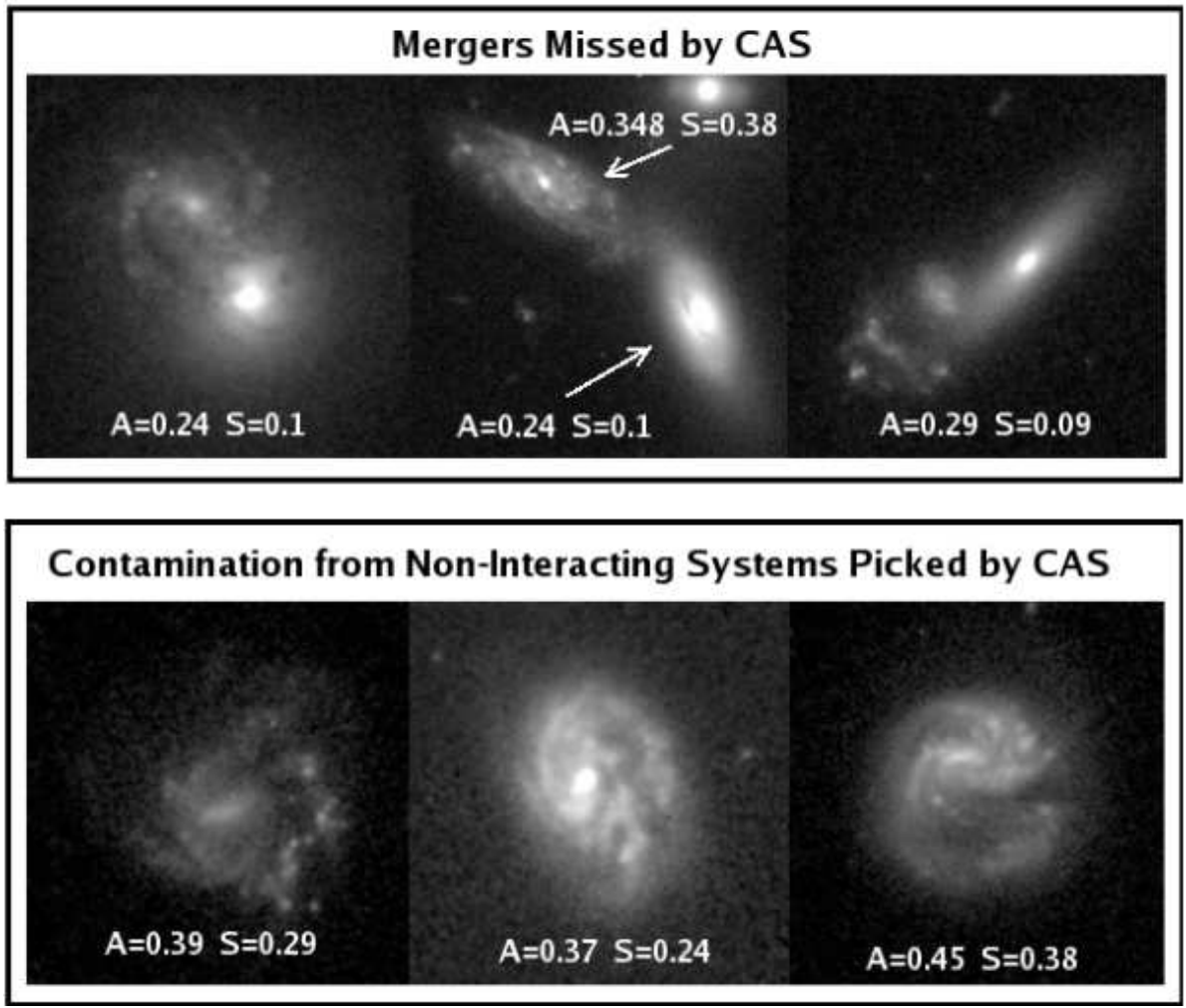}
\caption{ The top panel shows examples of mergers missed by the CAS
merger criterion ($A > S$ and $A > 0.35$). Features of these systems
include double nuclei, tidal bridges, and tidal debris.  The lower
panel show some non-interacting contaminants picked up by the CAS
merger criterion, due to small-scale asymmetries from SF, strong
patchy dust lanes, and the absence of a clear center, all leading to a
larger $A$ value.}
\label{fcas2}
\end{figure}

\clearpage
\begin{figure}
\epsscale{.850}
\plotone{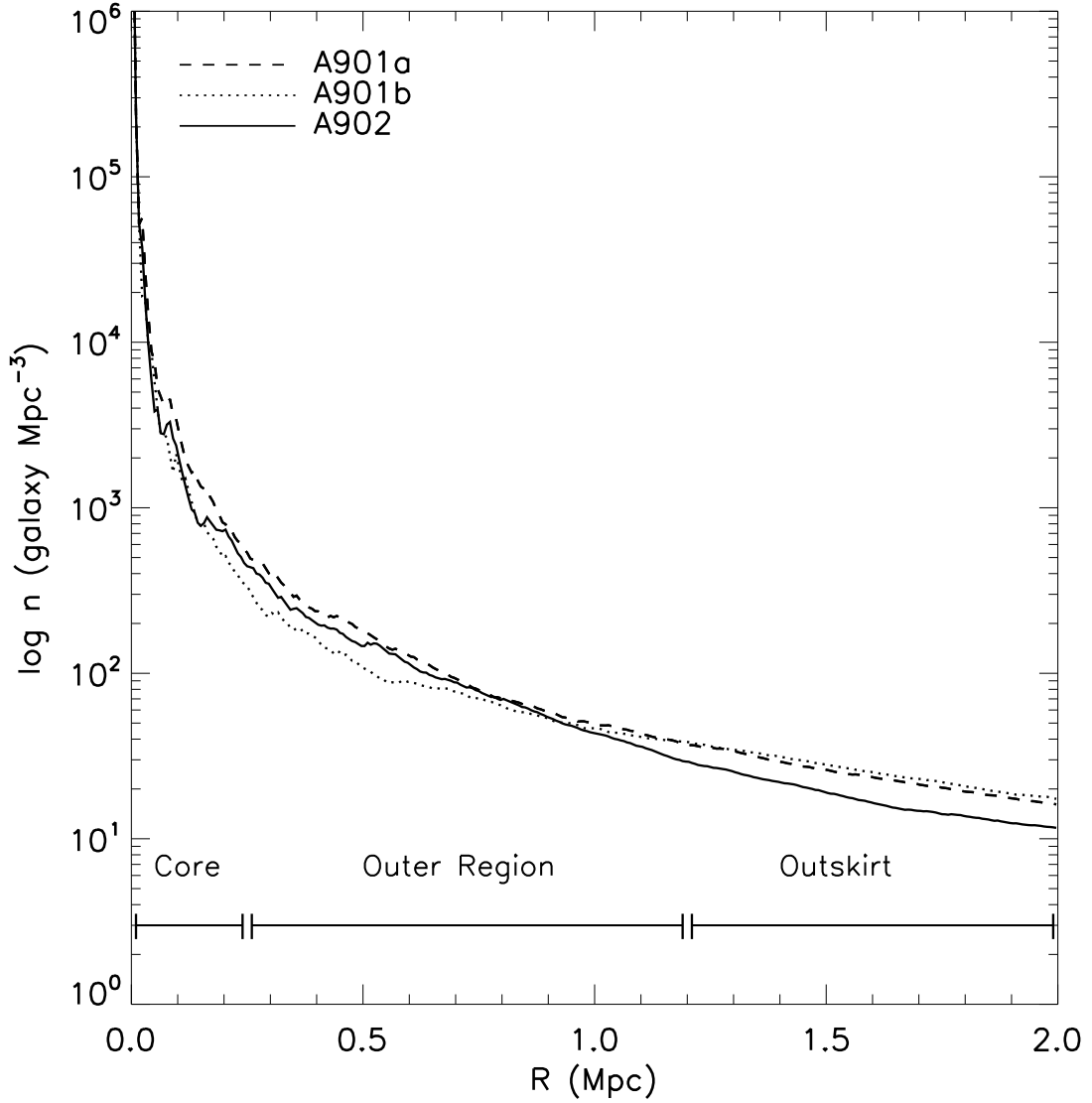}
\caption{The azimuthally averaged projected number density $n$ of
bright ($R_{\rm Vega} \le$~24) intermediate-mass ($M_{*} \geq10^{9}
M_{\sun}$) galaxies in the sample is shown as a function of
clustocentric radius which is shown for each cluster A901a/b and A902.
We consider the cluster core to be at $R\le$~0.25 Mpc, as this is the
region where the projected number density $n$ rises very steeply.  We
refer to the region which is located at 0.25 Mpc~$<R \le$~1.2 Mpc
between the cluster core and the cluster virial radius as the outer
region of the cluster. The region outside the virial radius (1.2
Mpc~$< R \le$~2.0 Mpc) is referred to as the outskirt region of the
cluster. The core, outer region, and outskirt region are labeled.}
\label{fnumd}
\end{figure}

\clearpage
\begin{figure}
\epsscale{.850}
\plotone{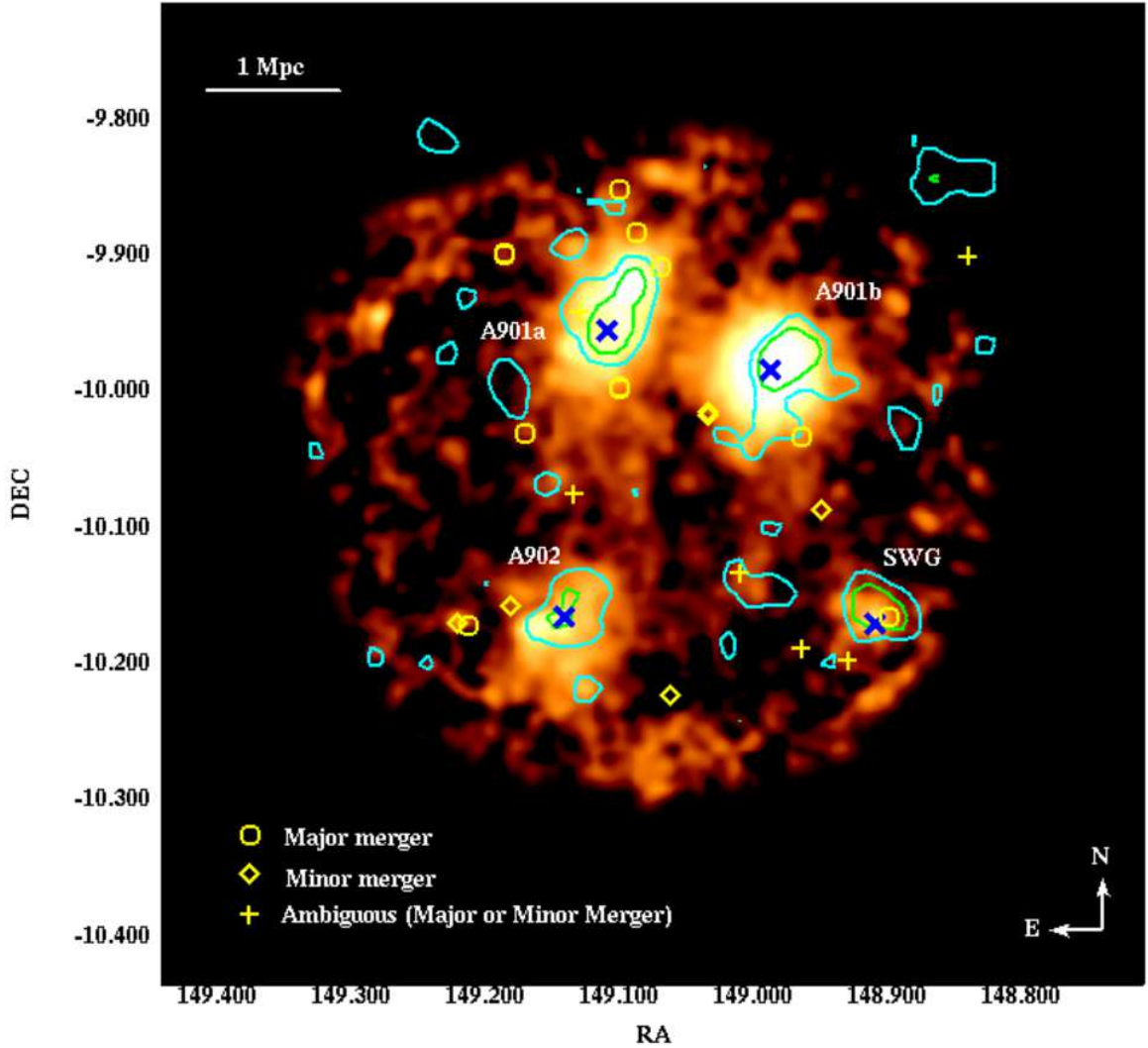}
\caption{ The distribution of visually-classified mergers (coded as
yellow diamonds, circles, and crosses) among the final sample of
bright ($R_{\rm Vega} \le$~24) intermediate-mass ($M_{*} \geq 10^{9}
M_{\sun}$ galaxies) is shown, overlaid on the ICM density map
(yellow-orange scale).  We only plot here the final sample of 20
distorted mergers, shown with starred identification numbers in
Table~\ref{tmer}.  Following Heymans \etal (2008), we show DM masses
in terms of the signal-to-noise of the weak lensing detection.  The
cyan and green contours enclose roughly 6h$^{-1}$10$^{13}M_{\sun}$ and
3.5h$^{-1}$10$^{13}M_{\sun}$ for A901a and A901b, and roughly
3h$^{-1}$10$^{13}M_{\sun}$ and 1.5h$^{-1}$10$^{13}M_{\sun}$ for the
lower mass A902 and South West Group (SWG).  The different symbols
represent major mergers (circles), minor mergers (diamonds), and
ambiguous major or minor merger cases (crosses), identified in
$\S$Section~\ref{smvc}. All mergers are located in the outer region and
outskirt of each cluster (0.25 Mpc~$< R \le$~2 Mpc).  }
\label{fdist1}
\end{figure}

\clearpage
\begin{figure}
\epsscale{0.8}
\plotone{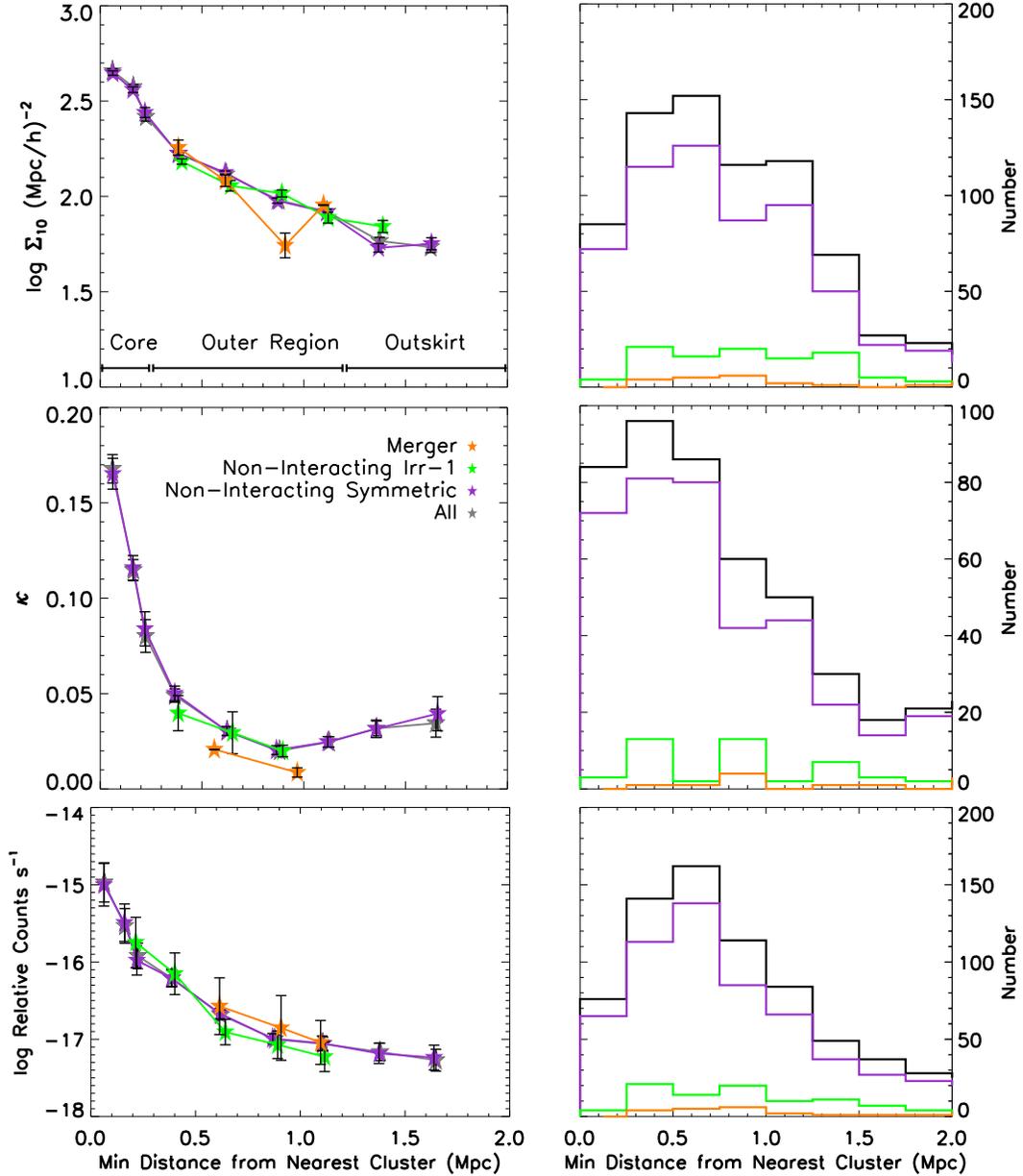}
\caption{ For mergers and non-interacting galaxies, the minimum
distance to the nearest cluster center (A901a, A901b, A902) is plotted
against the values of various local environmental parameters, such as
the local galaxy surface density ($\Sigma_{10}$) {\bf (top left)}, the
local DM mass surface density ($\kappa$) {\bf (middle left)}, and the
relative local ICM density {\bf (bottom left)}.  The panels on the
right--hand side show the number of mergers and non-interacting
galaxies which are found at different for each parameter.  Mergers lie
in the outer region and outskirt of the cluster (0.25 Mpc~$< R \le$2
Mpc) and are associated with low values of $\kappa$ and intermediate
values of $\Sigma_{10}$, and ICM density.  }
\label{fnsigk}
\end{figure}

\clearpage
\begin{figure}
\epsscale{0.9}
\plotone{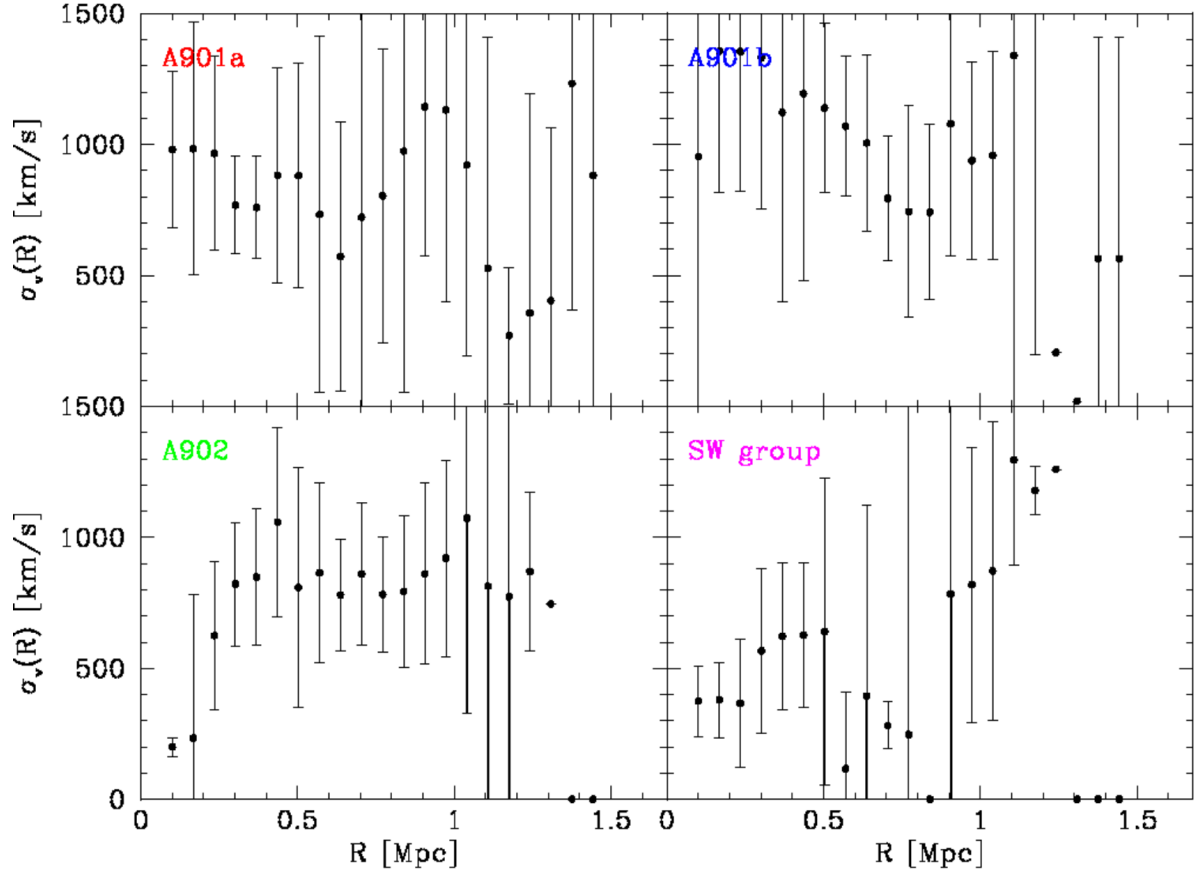}
\caption{Local velocity dispersion profiles for A901a, b and A902
clusters, and associated SWG from kinematic modeling using the
$\sim$420 2dF redshifts (Gray \etal in preparation).  The central galaxy
velocity dispersion within the cores ($R<0.25$~Mpc) of A901a,b and
A902 typically range from 700 to 1000 km s$^{-1}$.  Outside the
cluster core, in the outer region (0.25 Mpc~$< R \le$~1.2 Mpc), the
small number statistics leads to large error bars on the galaxy
velocity dispersion, making it not viable to determine whether it
remains high or drops.  }
\label{fcvdp}
\end{figure}

\clearpage
\begin{figure}
\epsscale{0.9}
\plotone{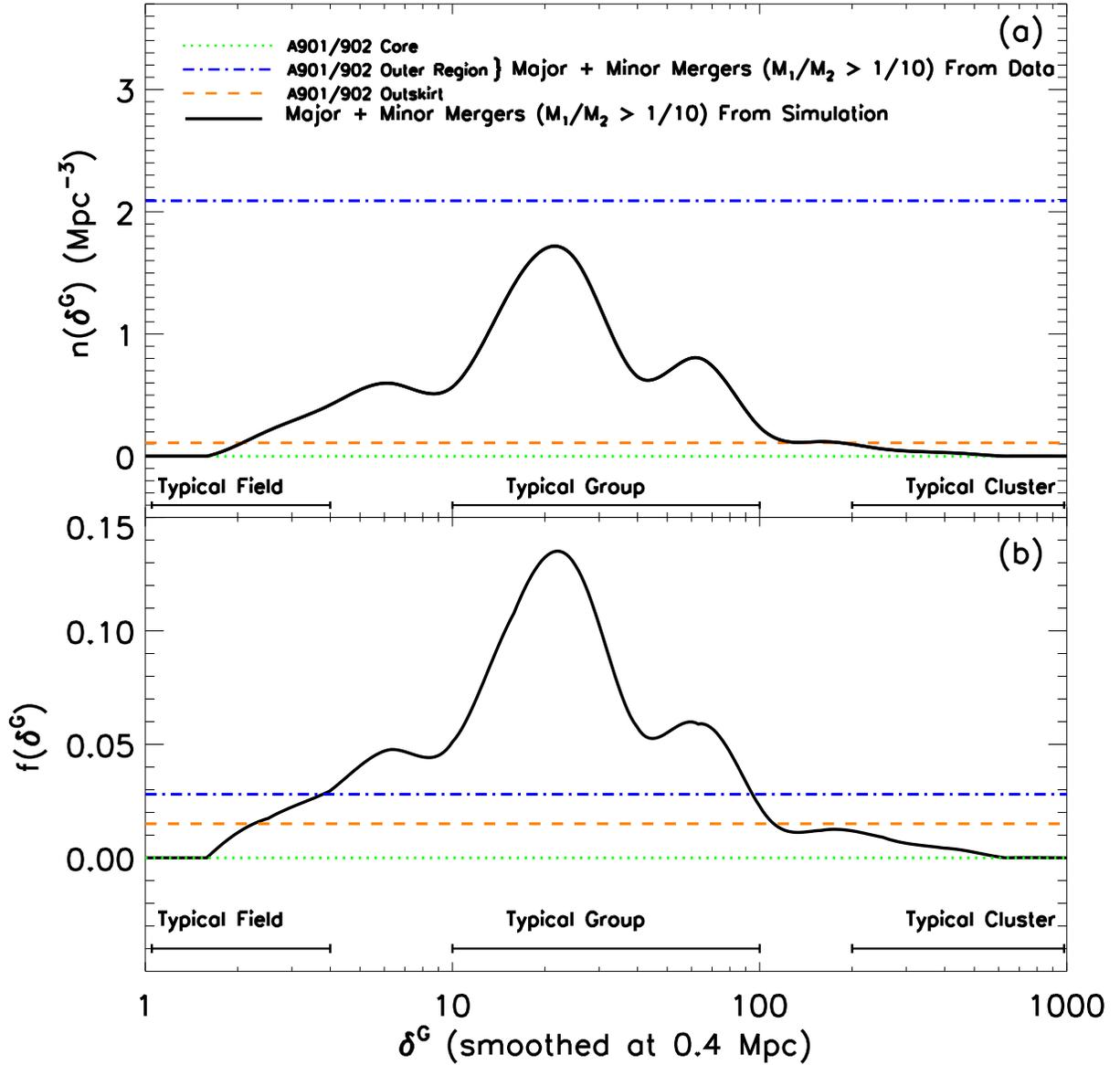}
\caption{ Comparison of the observed and theoretically
predicted number density of mergers {\bf (top panel)} and fraction of
mergers {\bf (lower panel)}, as detailed in $\S$Section~\ref{sinte3}.  The
solid curve shows the {\it predicted model} number density of mergers
($n(\delta^{\rm G})$) and the fraction of mergers ($f(\delta^{\rm
G})$) as a function of local overdensity ($\delta^{\rm G}$), in the
$N$-body simulations of the STAGES A901/902 supercluster (van Kampen
\etal 2009, in preparation).  The local overdensity is calculated by
smoothing the density of DM halos with a Gaussian of width
0.4 Mpc to takeout the effect of individual galaxies.  Typical values
of $\delta^{\rm G}$ are $\sim$10-100 for group overdensities,
$\sim$200 at the cluster virial radius, and $\gtrsim$~1000 in core of
rich clusters.  In the models, as field and group galaxies fall into a
cluster along filaments, the bulk flow enhances the galaxy density and
causes galaxies to have small relative velocities, thus leading to a
high probability for mergers at typical group overdensities.  Closer
to the cluster core, model galaxies show large random motions,
producing a sharp drop in the probability of mergers.  The three dashed
lines show the {\it observed} number density ($n_{\rm merge}$) and
fraction ($f_{\rm merge}$) of mergers in three different regions of the
A901/902 clusters: the core, the outer region, and the outskirt.  A
stellar mass cut of $M>$ 10$^{9} M_{\sun}$ is applied to both model
and data. The points at which the three dashed lines cross or approach the
solid curve tell us the typical overdensities at which we expect to
find such merger fraction or merger number densities in the
simulations.  It can be seen that the low merger density seen in the
core region (green dashed line) of the cluster correspond to those
expected at typical cluster core overdensities.  On the other hand,
the larger merger fraction we observe between the cluster core and
outer region (0.25 Mpc~$< R \le$~2 Mpc; blue dashed line) is close to
those seen in typical {\it group overdensities}, in agreement with the
scenario of cluster growth via accretion of group and/or field
galaxies.  }
\label{felco1}
\end{figure}

\clearpage
\begin{figure}
\epsscale{.90}
\plotone{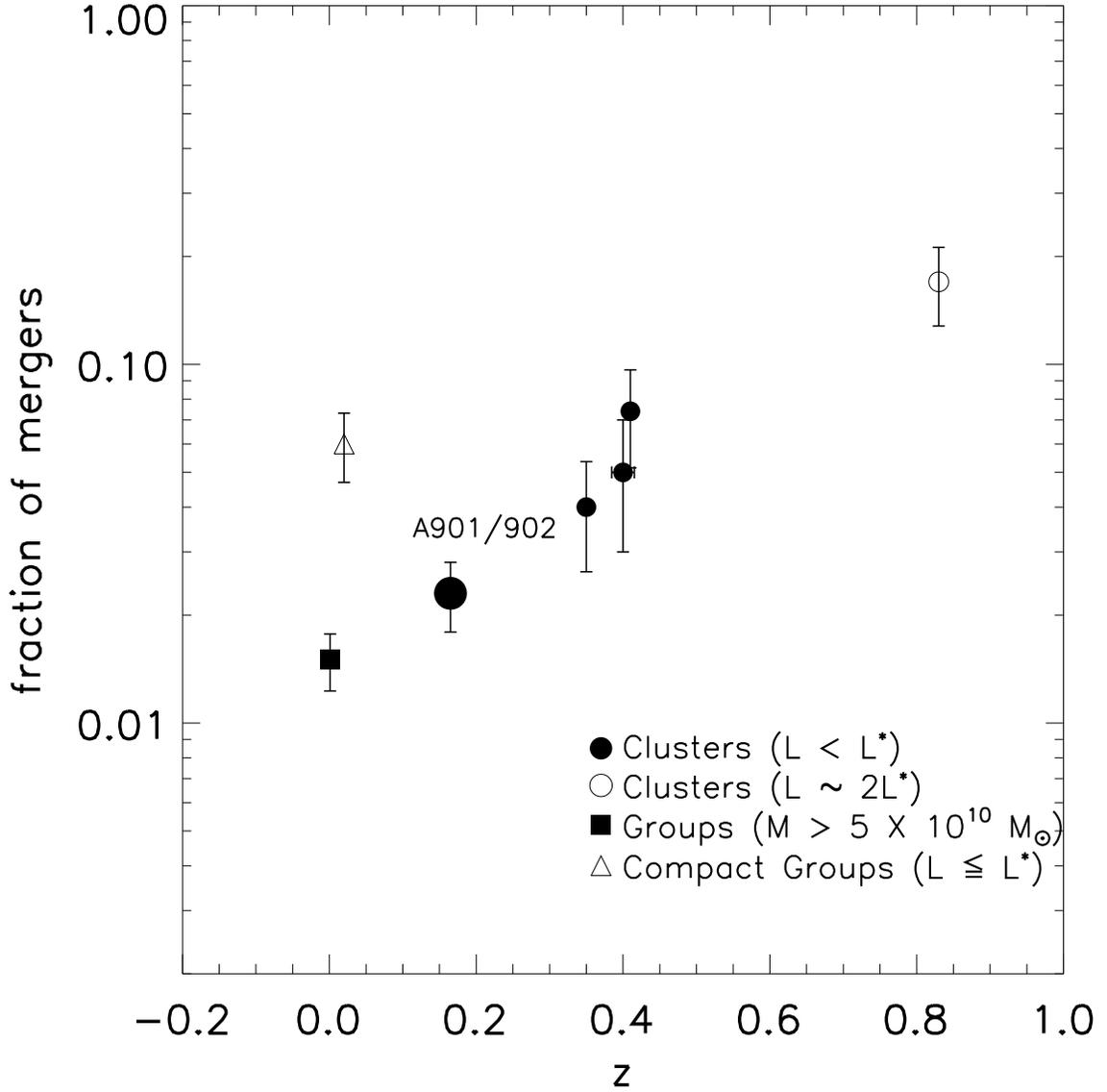}
\caption{Comparisons of our observed fraction of mergers among
intermediate mass systems in the A901/902 clusters (large filled solid
circle) with the results from other studies of clusters and groups.
Filled circles are for cluster data points of intermediate luminosity
($L < L^{*}$ and/or $M = 10^{9}$ to a few $\times 10^{10} M_{\sun}$),
and include the following studies in order of increasing redshift:
this study of A901/902 (most systems have $M_{V}\sim$~-19 to~-22 and
$M_{*} = 10^{9}$ to a few $\times 10^{10} M_{\sun}$); Couch \etal
(1998; $L < 2L^{*}$), Oemler \etal (1997; lower limit of $f$ for
$M_{V} < -19$), Dressler \etal (1994; $M_{V} < -18.5$). The van Dokkum
\etal (1999) point at $z=0.83$ (open circle) is for a cluster sample
of luminous ($M_{B}\sim-22$ and $L\sim2L^{*}$) galaxies.  The group
point (filled square) is from McIntosh \etal (2008) for galaxy masses
above $5 \times 10^{10} M_{\sun}$.  The Hickson compact group point
(open triangle), from Zepf \etal (1993), is for galaxies of
luminosities $L \leq L^{*}$.  }
\label{fcompa}
\end{figure}

\clearpage
\begin{figure}
\epsscale{.8}
\plotone{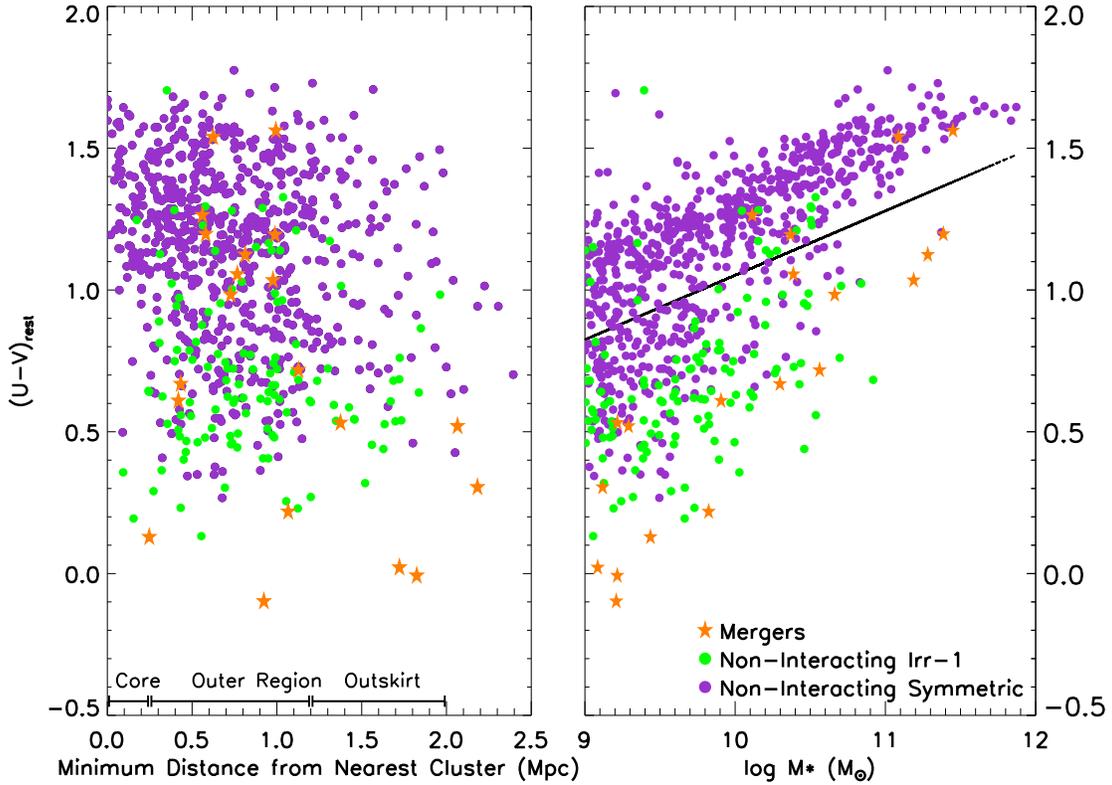}
\caption{ {\bf Left:} the rest-frame $U-V$ color is plotted against
the minimum distance from the nearest cluster center, for systems of
different visual classes (Merger, Non-interacting Irr-1, and
Non-interacting Symmetric).  We only consider here the 20 distorted
mergers in Table~\ref{tmer}, split into 13 mergers of type 1, 3
mergers of type 2a, and 4 mergers of type 2b.  For mergers of type 2b,
which are resolved into two galaxies with separate COMBO-17 colors, we
plot the average $U-V$ color of the galaxies in the pair.  {\bf
Right:} the rest-frame $U-V$ color is plotted against stellar masses .
The black solid lines separate the blue cloud and red sample.  For $M
\ge 10^{9} M_{\sun}$ systems, we find that $80\% \pm 18\%$ (16/20) of
mergers lie on the blue cloud, compared to (294/866) or $34\%\pm 7\%$ of
the non-interacting galaxies. Thus, mergers and interacting galaxies
are preferentially blue compared to non-interacting galaxies. }
\label{fcmd1}
\end{figure}

\clearpage
\begin{figure}
\epsscale{.90}
\plotone{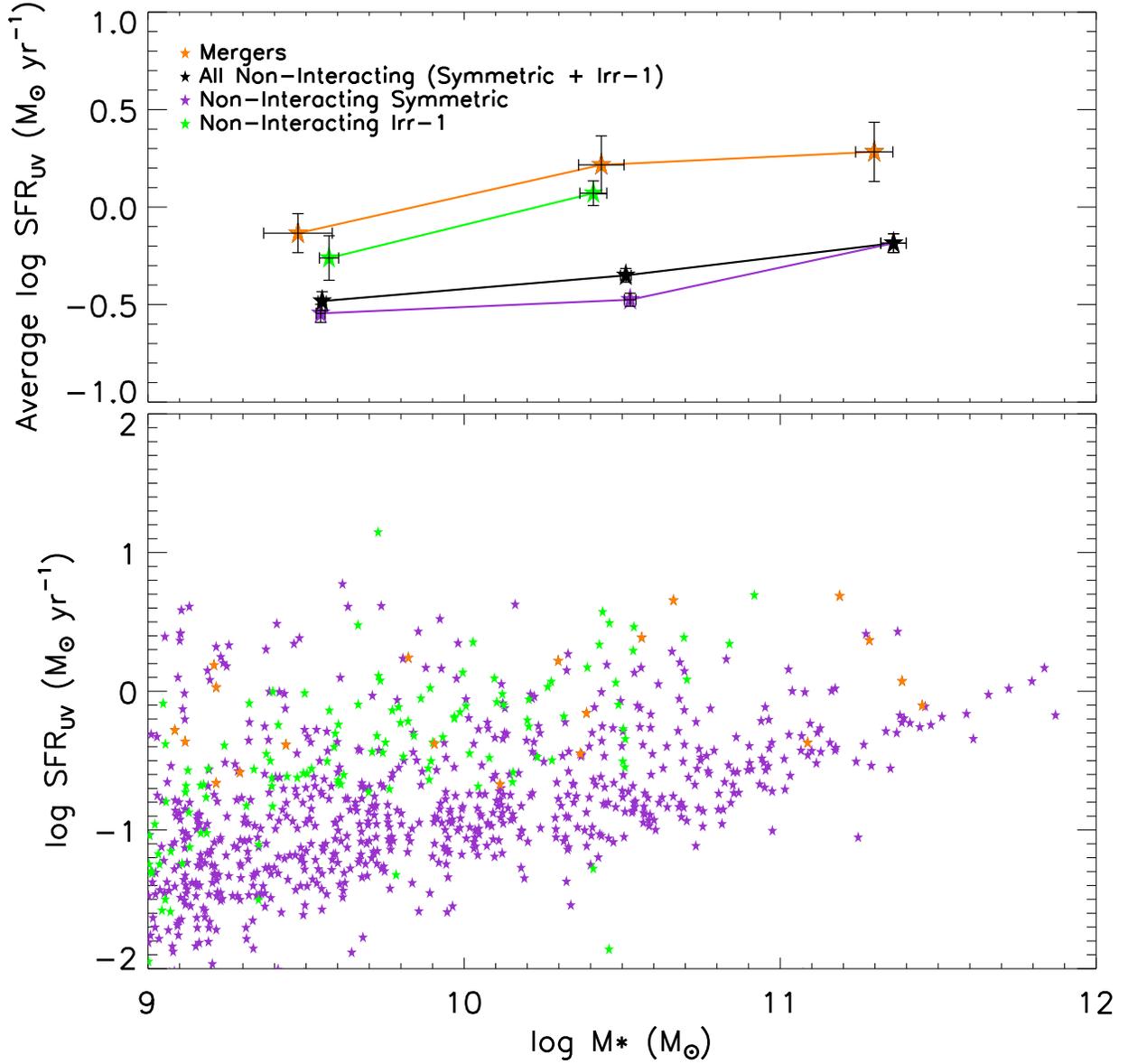}
\caption{ {\bf Bottom panel:} The UV-based SFR is plotted against
stellar mass for the sample of 886 bright intermediate-mass ($M_{*}
\geq 10^{9} M_{\sun}$) visually classifiable systems.  Systems are
coded according to their visual classes: Merger, Non-interacting
Irr-1, and Non-interacting Symmetric.  We only consider here the 20
distorted mergers in Table~\ref{tmer}.  {\bf Top panel:} The average
UV-based SFR is plotted against stellar mass for galaxies of different
visual classes.  For the few mergers galaxies present (orange line),
the average SFR$_{\rm UV}$ is enhanced by at most an average factor of
$\sim$2 compared to the Non-interacting Symmetric galaxies (purple
line) and to all Non-interacting galaxies (i.e Symmetric + Irr-1;
black line). No enhancement is seen with respect to the
Non-interacting Irr-1 galaxies (green line).  }
\label{fsfruv}
 \end{figure}

\clearpage
\begin{figure}
\epsscale{.95}
\plotone{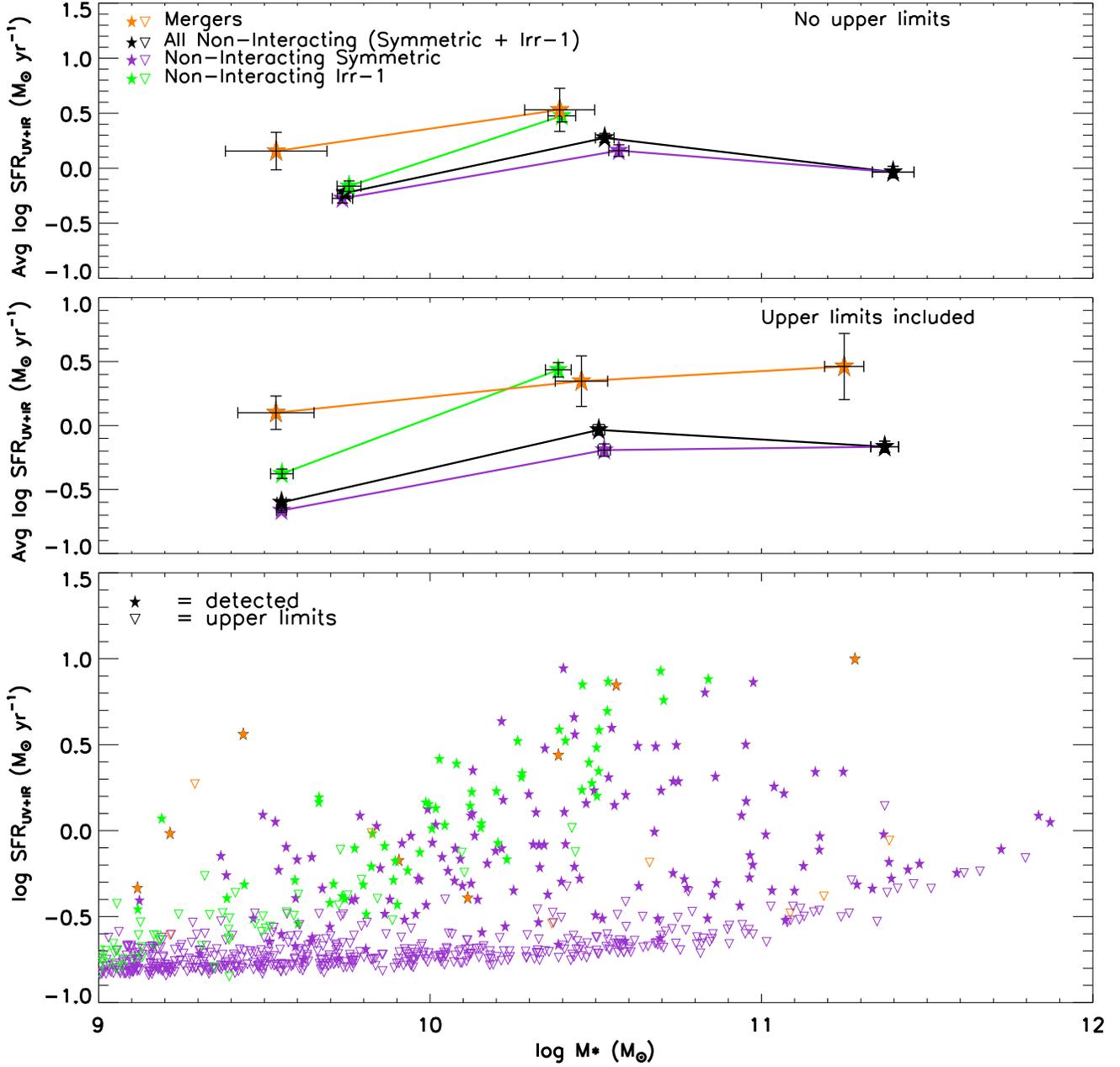}
\caption{ {\bf Bottom panel:} the UV+IR-based SFR is plotted against
stellar mass for the sample of bright intermediate-mass ($M_{*} \geq
10^{9} M_{\sun}$) systems.  Galaxies are coded according to their
visual classes: Merger, Non-interacting Irr-1, and Non-interacting
Symmetric.  We only consider here the 20 distorted mergers in
Table~\ref{tmer}.  For galaxies that were observed and detected at
24$\micron$, the UV+IR-based SFR is plotted as stars.  For galaxies
that were observed but undetected at 24$\micron$, we use the detection
limit as an upper limit for the UV+IR-based SFR, and plot this limit
as inverted triangles.  {\bf Middle panel:} the average UV+IR-based
SFR is plotted against stellar mass for galaxies of different visual
classes. We include galaxies with a 24$\micron$ detection, as well as
those with only upper limits on the 24$\micron$ flux.  {\bf Top
panel:} same as middle panel, except that we only include galaxies
with a 24$\micron$ detection and exclude those with upper limits.  In
both top and middle panels, we find that for the few mergers present
(orange line), the average SFR$_{\rm UV+IR}$ is typically enhanced by
only a factor of $\sim$~1.5 compared to the Non-interacting Symmetric
galaxies (purple line) and to all Non-interacting galaxies (i.e
Symmetric + Irr-1; black line). }
\label{fsfrtot}
 \end{figure}

\end{document}